\documentstyle[12pt]{article}
\def\fileversion{v1.13}%
\def\filedate{6.5.93}%
\edef\epsfigRestoreAt{\catcode`@=\number\catcode`@\relax}%
\catcode`\@=11\relax
\immediate\write16{Document style option `epsfig', \fileversion\space
<\filedate> (edited by SPQR)}%
\newcount\EPS@Height
\newcount\EPS@Width
\newcount\EPS@xscale
\newcount\EPS@yscale
\def\psfigdriver#1{%
  \bgroup\edef\next{\def\noexpand\tempa{#1}}%
    \uppercase\expandafter{\next}%
    \def\LN{DVITOLN03}%
    \def\DVItoPS{DVITOPS}%
    \def\DVIPS{DVIPS}%
    \def\emTeX{EMTEX}%
    \def\OzTeX{OZTEX}%
    \def\Textures{TEXTURES}%
    \global\chardef\fig@driver=0
    \ifx\tempa\LN
        \global\chardef\fig@driver=0\fi
    \ifx\tempa\DVItoPS
        \global\chardef\fig@driver=1\fi
    \ifx\tempa\DVIPS
        \global\chardef\fig@driver=2\fi
    \ifx\tempa\emTeX
        \global\chardef\fig@driver=3\fi
    \ifx\tempa\OzTeX
        \global\chardef\fig@driver=4\fi
    \ifx\tempa\Textures
        \global\chardef\fig@driver=5\fi
  \egroup
\def\psfig@start{}%
\def\psfig@end{}%
\def\epsfig@gofer{}%
\ifcase\fig@driver
\typeout{WARNING! ****
 no specials for LN03 psfig}%
\or 
\def\psfig@start{}%
\def\psfig@end{\special{dvitops: import \@p@sfilefinal \space
\@p@swidth sp \space \@p@sheight sp \space fill}%
\if@clip \typeout{Clipping not supported}\fi
\if@angle \typeout{Rotating not supported}\fi
}%
\let\epsfig@gofer\psfig@end
\or 
\def\psfig@start{\special{ps::[begin]  \@p@swidth \space \@p@sheight \space%
        \@p@sbbllx \space \@p@sbblly \space%
        \@p@sbburx \space \@p@sbbury \space%
        startTexFig \space }%
        \if@angle
                \special {ps:: \@p@sangle \space rotate \space}
        \fi
        \if@clip
                \if@verbose
                        \typeout{(clipped to BB) }%
                \fi
                \special{ps:: doclip \space }%
        \fi
        \special{ps: plotfile \@p@sfilefinal \space }%
        \special{ps::[end] endTexFig \space }%
}%
\def\psfig@end{}%
\def\epsfig@gofer{\if@clip
                        \if@verbose
                           \typeout{(clipped to BB)}%
                        \fi
                        \epsfclipon
                  \fi
                  \epsfsetgraph{\@p@sfilefinal}%
}%
\or 
\typeout{WARNING. You must have a .bb info file with the Bounding Box
  of the pcx file}%
\def\psfig@start{}%
\def\psfig@end{\typeout{pcx import of \@p@sfilefinal}%
\if@clip \typeout{Clipping not supported}\fi
\if@angle \typeout{Rotating not supported}\fi
\raisebox{\@p@srheight true sp}{\special{em: graph \@p@sfilefinal}}}%
\def\epsfig@gofer{}%
\or 
\def\psfig@start{}%
\def\psfig@end{%
\EPS@Width\@p@swidth
\EPS@Height\@p@sheight
\divide\EPS@Width by 65781  
\divide\EPS@Height by 65781
\special{epsf=\@p@sfilefinal
\space
width=\the\EPS@Width
\space
height=\the\EPS@Height
}%
\if@clip \typeout{Clipping not supported}\fi
\if@angle \typeout{Rotating not supported}\fi
}%
\let\epsfig@gofer\psfig@end
\or 
\def\psfig@end{\if@clip
                        \if@verbose
                           \typeout{(clipped to BB)}%
                        \fi
                        \epsfclipon
                  \fi
\special{illustration \@p@sfilefinal\space scaled \the\EPS@xscale}%
}%
\def\psfig@start{}%
\let\epsfig\psfig
\else
\typeout{WARNING. *** unknown  driver - no psfig}%
\fi
}%
\newdimen\ps@dimcent
\ifx\undefined\fbox
\newdimen\fboxrule
\newdimen\fboxsep
\newdimen\ps@tempdima
\newbox\ps@tempboxa
\long\def\fbox#1{\leavevmode\setbox\ps@tempboxa\hbox{#1}\ps@tempdima\fboxrule
    \advance\ps@tempdima \fboxsep \advance\ps@tempdima \dp\ps@tempboxa
   \hbox{\lower \ps@tempdima\hbox
  {\vbox{\hrule height \fboxrule
          \hbox{\vrule width \fboxrule \hskip\fboxsep
          \vbox{\vskip\fboxsep \box\ps@tempboxa\vskip\fboxsep}\hskip
                 \fboxsep\vrule width \fboxrule}%
                 \hrule height \fboxrule}}}}%
\fi
\ifx\@ifundefined\undefined
\long\def\@ifundefined#1#2#3{\expandafter\ifx\csname
  #1\endcsname\relax#2\else#3\fi}%
\fi
\@ifundefined{typeout}%
{\gdef\typeout#1{\immediate\write\sixt@@n{#1}}}%
{\relax}%
\@ifundefined{epsfig}{}{\typeout{EPSFIG --- already loaded}\endinput}%
\@ifundefined{epsfbox}{\input epsf}{}%
\ifx\undefined\@latexerr
        \newlinechar`\^^J
        \def\@spaces{\space\space\space\space}%
        \def\@latexerr#1#2{%
        \edef\@tempc{#2}\expandafter\errhelp\expandafter{\@tempc}%
        \typeout{Error. \space see a manual for explanation.^^J
         \space\@spaces\@spaces\@spaces Type \space H <return> \space for
         immediate help.}\errmessage{#1}}%
\fi
\def\@whattodo{You tried to include a PostScript figure which
cannot be found^^JIf you press return to carry on anyway,^^J
The failed name will be printed in place of the figure.^^J
or type X to quit}%
\def\@whattodobb{You tried to include a PostScript figure which
has no^^Jbounding box, and you supplied none.^^J
If you press return to carry on anyway,^^J
The failed name will be printed in place of the figure.^^J
or type X to quit}%
\def\@nnil{\@nil}%
\def\@empty{}%
\def\@psdonoop#1\@@#2#3{}%
\def\@psdo#1:=#2\do#3{\edef\@psdotmp{#2}\ifx\@psdotmp\@empty \else
    \expandafter\@psdoloop#2,\@nil,\@nil\@@#1{#3}\fi}%
\def\@psdoloop#1,#2,#3\@@#4#5{\def#4{#1}\ifx #4\@nnil \else
       #5\def#4{#2}\ifx #4\@nnil \else#5\@ipsdoloop #3\@@#4{#5}\fi\fi}%
\def\@ipsdoloop#1,#2\@@#3#4{\def#3{#1}\ifx #3\@nnil
       \let\@nextwhile=\@psdonoop \else
      #4\relax\let\@nextwhile=\@ipsdoloop\fi\@nextwhile#2\@@#3{#4}}%
\def\@tpsdo#1:=#2\do#3{\xdef\@psdotmp{#2}\ifx\@psdotmp\@empty \else
    \@tpsdoloop#2\@nil\@nil\@@#1{#3}\fi}%
\def\@tpsdoloop#1#2\@@#3#4{\def#3{#1}\ifx #3\@nnil
       \let\@nextwhile=\@psdonoop \else
      #4\relax\let\@nextwhile=\@tpsdoloop\fi\@nextwhile#2\@@#3{#4}}%
\long\def\epsfaux#1#2:#3\\{\ifx#1\epsfpercent
   \def\testit{#2}\ifx\testit\epsfbblit
        \@atendfalse
        \epsf@atend #3 . \\%
        \if@atend
           \if@verbose
                \typeout{epsfig: found `(atend)'; continuing search}%
           \fi
        \else
                \epsfgrab #3 . . . \\%
                \epsffileokfalse\global\no@bbfalse
                \global\epsfbbfoundtrue
        \fi
   \fi\fi}%
\def\epsf@atendlit{(atend)}
\def\epsf@atend #1 #2 #3\\{%
   \def\epsf@tmp{#1}\ifx\epsf@tmp\empty
      \epsf@atend #2 #3 .\\\else
   \ifx\epsf@tmp\epsf@atendlit\@atendtrue\fi\fi}%

\chardef\trig@letter = 11
\chardef\other = 12

\newif\ifdebug 
\newif\ifc@mpute 
\newif\if@atend
\c@mputetrue 

\let\then = \relax
\def\r@dian{pt }%
\let\r@dians = \r@dian
\let\dimensionless@nit = \r@dian
\let\dimensionless@nits = \dimensionless@nit
\def\internal@nit{sp }%
\let\internal@nits = \internal@nit
\newif\ifstillc@nverging
\def \Mess@ge #1{\ifdebug \then \message {#1} \fi}%

{ 
        \catcode `\@ = \trig@letter
        \gdef \nodimen {\expandafter \n@dimen \the \dimen}%
        \gdef \term #1 #2 #3%
               {\edef \t@ {\the #1}
                \edef \t@@ {\expandafter \n@dimen \the #2\r@dian}%
                \t@rm {\t@} {\t@@} {#3}%
               }%
        \gdef \t@rm #1 #2 #3%
               {{%
                \count 0 = 0
                \dimen 0 = 1 \dimensionless@nit
                \dimen 2 = #2\relax
                \Mess@ge {Calculating term #1 of \nodimen 2}%
                \loop
                \ifnum  \count 0 < #1
                \then   \advance \count 0 by 1
                        \Mess@ge {Iteration \the \count 0 \space}%
                        \Multiply \dimen 0 by {\dimen 2}%
                        \Mess@ge {After multiplication, term = \nodimen 0}%
                        \Divide \dimen 0 by {\count 0}%
                        \Mess@ge {After division, term = \nodimen 0}%
                \repeat
                \Mess@ge {Final value for term #1 of
                                \nodimen 2 \space is \nodimen 0}%
                \xdef \Term {#3 = \nodimen 0 \r@dians}%
                \aftergroup \Term
               }}%
        \catcode `\p = \other
        \catcode `\t = \other
        \gdef \n@dimen #1pt{#1} 
}%

\def \Divide #1by #2{\divide #1 by #2} 

\def \Multiply #1by #2
       {{
        \count 0 = #1\relax
        \count 2 = #2\relax
        \count 4 = 65536
        \Mess@ge {Before scaling, count 0 = \the \count 0 \space and
                        count 2 = \the \count 2}%
        \ifnum  \count 0 > 32767 
        \then   \divide \count 0 by 4
                \divide \count 4 by 4
        \else   \ifnum  \count 0 < -32767
                \then   \divide \count 0 by 4
                        \divide \count 4 by 4
                \else
                \fi
        \fi
        \ifnum  \count 2 > 32767 
        \then   \divide \count 2 by 4
                \divide \count 4 by 4
        \else   \ifnum  \count 2 < -32767
                \then   \divide \count 2 by 4
                        \divide \count 4 by 4
                \else
                \fi
        \fi
        \multiply \count 0 by \count 2
        \divide \count 0 by \count 4
        \xdef \product {#1 = \the \count 0 \internal@nits}%
        \aftergroup \product
       }}%

\def\r@duce{\ifdim\dimen0 > 90\r@dian \then   
                \multiply\dimen0 by -1
                \advance\dimen0 by 180\r@dian
                \r@duce
            \else \ifdim\dimen0 < -90\r@dian \then  
                \advance\dimen0 by 360\r@dian
                \r@duce
                \fi
            \fi}%

\def\Sine#1%
       {{%
        \dimen 0 = #1 \r@dian
        \r@duce
        \ifdim\dimen0 = -90\r@dian \then
           \dimen4 = -1\r@dian
           \c@mputefalse
        \fi
        \ifdim\dimen0 = 90\r@dian \then
           \dimen4 = 1\r@dian
           \c@mputefalse
        \fi
        \ifdim\dimen0 = 0\r@dian \then
           \dimen4 = 0\r@dian
           \c@mputefalse
        \fi
        \ifc@mpute \then
                \divide\dimen0 by 180
                \dimen0=3.141592654\dimen0
                \dimen 2 = 3.1415926535897963\r@dian 
                \divide\dimen 2 by 2 
                \Mess@ge {Sin: calculating Sin of \nodimen 0}%
                \count 0 = 1 
                \dimen 2 = 1 \r@dian 
                \dimen 4 = 0 \r@dian 
                \loop
                        \ifnum  \dimen 2 = 0 
                        \then   \stillc@nvergingfalse
                        \else   \stillc@nvergingtrue
                        \fi
                        \ifstillc@nverging 
                        \then   \term {\count 0} {\dimen 0} {\dimen 2}%
                                \advance \count 0 by 2
                                \count 2 = \count 0
                                \divide \count 2 by 2
                                \ifodd  \count 2 
                                \then   \advance \dimen 4 by \dimen 2
                                \else   \advance \dimen 4 by -\dimen 2
                                \fi
                \repeat
        \fi
                        \xdef \sine {\nodimen 4}%
       }}%

\def\Cosine#1{\ifx\sine\UnDefined\edef\Savesine{\relax}\else
                             \edef\Savesine{\sine}\fi
        {\dimen0=#1\r@dian\multiply\dimen0 by -1
         \advance\dimen0 by 90\r@dian
         \Sine{\nodimen 0}%
         \xdef\cosine{\sine}%
         \xdef\sine{\Savesine}}}
\def\psdraft{\def\@psdraft{0}}%
\def\psfull{\def\@psdraft{1}}%
\psfull
\newif\if@scalefirst
\def\psscalefirst{\@scalefirsttrue}%
\def\psrotatefirst{\@scalefirstfalse}%
\psrotatefirst
\newif\if@draftbox
\def\psnodraftbox{\@draftboxfalse}%
\@draftboxtrue
\newif\if@noisy
\@noisyfalse
\newif\ifno@bb
\newif\if@bbllx
\newif\if@bblly
\newif\if@bburx
\newif\if@bbury
\newif\if@height
\newif\if@width
\newif\if@rheight
\newif\if@rwidth
\newif\if@angle
\newif\if@clip
\newif\if@verbose
\newif\if@prologfile
\def\@p@@sprolog#1{\@prologfiletrue\def\@prologfileval{#1}}%
\def\@p@@sclip#1{\@cliptrue}%
\newif\ifepsfig@dos  
\def\epsfigdos{\epsfig@dostrue}%
\epsfig@dosfalse
\newif\ifuse@psfig
\def\ParseName#1{\expandafter\@Parse#1}%
\def\@Parse#1.#2:{\gdef\BaseName{#1}\gdef\FileType{#2}}%

\def\@p@@sfile#1{%
\ifepsfig@dos
   \ParseName{#1:}%
\else
   \gdef\BaseName{#1}\gdef\FileType{}%
\fi
\def\@p@sfile{NO FILE: #1}%
\def\@p@sfilefinal{NO FILE: #1}%
        \openin1=#1
        \ifeof1\closein1
                \openin1=\BaseName.bb
                        \ifeof1\closein1
                                \if@bbllx\if@bblly\if@bburx\if@bbury
                                        \def\@p@sfile{#1}%
                                        \def\@p@sfilefinal{#1}%
                                        \fi\fi\fi
                                \else
                                        \@latexerr{ERROR.
PostScript file #1 not found}\@whattodo
                                        \@p@@sbbllx{100bp}%
                                        \@p@@sbblly{100bp}%
                                        \@p@@sbburx{200bp}%
                                        \@p@@sbbury{200bp}%
                                        \psdraft
                                \fi
                        \else
                                \closein1%
                                \edef\@p@sfile{\BaseName.bb}%
                                \typeout{using BB from \@p@sfile}%
                                \ifnum\fig@driver=3
                                  \edef\@p@sfilefinal{\BaseName.pcx}%
                                \else
                                 \ifepsfig@dos
                                 \edef\@p@sfilefinal{"`uncompress
                                   < \BaseName.Z"}%
                                \else
                                \edef\@p@sfilefinal{"`zcat `texfind
                                  #1.Z`"}%
                                \fi
                                \fi
                        \fi
        \else\closein1
                    \edef\@p@sfile{#1}%
                    \edef\@p@sfilefinal{#1}%
        \fi%
}%
\let\@p@@sfigure\@p@@sfile
\def\@p@@sbbllx#1{%
				            \@bbllxtrue
                \ps@dimcent=#1
                \edef\@p@sbbllx{\number\ps@dimcent}%
                \divide\ps@dimcent by65536
                \global\edef\epsfllx{\number\ps@dimcent}%
}%
\def\@p@@sbblly#1{%
                \@bbllytrue
                \ps@dimcent=#1
                \edef\@p@sbblly{\number\ps@dimcent}%
                \divide\ps@dimcent by65536
                \global\edef\epsflly{\number\ps@dimcent}%
}%
\def\@p@@sbburx#1{%
                \@bburxtrue
                \ps@dimcent=#1
                \edef\@p@sbburx{\number\ps@dimcent}%
                \divide\ps@dimcent by65536
                \global\edef\epsfurx{\number\ps@dimcent}%
}%
\def\@p@@sbbury#1{%
                \@bburytrue
                \ps@dimcent=#1
                \edef\@p@sbbury{\number\ps@dimcent}%
                \divide\ps@dimcent by65536
                \global\edef\epsfury{\number\ps@dimcent}%
}%
\def\@p@@sheight#1{%
                \@heighttrue
                \global\epsfysize=#1
                \ps@dimcent=#1
                \edef\@p@sheight{\number\ps@dimcent}%
}%
\def\@p@@swidth#1{%
                \@widthtrue
                \global\epsfxsize=#1
                \ps@dimcent=#1
                \edef\@p@swidth{\number\ps@dimcent}%
}%
\def\@p@@srheight#1{%
                \@rheighttrue\use@psfigtrue
                \ps@dimcent=#1
                \edef\@p@srheight{\number\ps@dimcent}%
}%
\def\@p@@srwidth#1{%
                \@rwidthtrue\use@psfigtrue
                \ps@dimcent=#1
                \edef\@p@srwidth{\number\ps@dimcent}%
}%
\def\@p@@sangle#1{%
                \use@psfigtrue
                \@angletrue
                \edef\@p@sangle{#1}%
}%
\def\@p@@ssilent#1{%
                \@verbosefalse
}%
\def\@p@@snoisy#1{%
                \@verbosetrue
}%
\def\@cs@name#1{\csname #1\endcsname}%
\def\@setparms#1=#2,{\@cs@name{@p@@s#1}{#2}}%
\def\ps@init@parms{%
                \@bbllxfalse \@bbllyfalse
                \@bburxfalse \@bburyfalse
                \@heightfalse \@widthfalse
                \@rheightfalse \@rwidthfalse
                \def\@p@sbbllx{}\def\@p@sbblly{}%
                \def\@p@sbburx{}\def\@p@sbbury{}%
                \def\@p@sheight{}\def\@p@swidth{}%
                \def\@p@srheight{}\def\@p@srwidth{}%
                \def\@p@sangle{0}%
                \def\@p@sfile{}%
                \use@psfigfalse
                \@prologfilefalse
                \def\@sc{}%
                \if@noisy
                        \@verbosetrue
                \else
                        \@verbosefalse
                \fi
                \@clipfalse
}%
\def\parse@ps@parms#1{%
                \@psdo\@psfiga:=#1\do
                   {\expandafter\@setparms\@psfiga,}%
\if@prologfile
\fi
}%
\def\bb@missing{%
        \if@verbose
            \typeout{psfig: searching \@p@sfile \space  for bounding box}%
        \fi
        \epsfgetbb{\@p@sfile}%
        \ifepsfbbfound
            \ps@dimcent=\epsfllx bp\edef\@p@sbbllx{\number\ps@dimcent}%
            \ps@dimcent=\epsflly bp\edef\@p@sbblly{\number\ps@dimcent}%
            \ps@dimcent=\epsfurx bp\edef\@p@sbburx{\number\ps@dimcent}%
            \ps@dimcent=\epsfury bp\edef\@p@sbbury{\number\ps@dimcent}%
        \else
            \epsfbbfoundfalse
        \fi
}
\newdimen\p@intvaluex
\newdimen\p@intvaluey
\def\rotate@#1#2{{\dimen0=#1 sp\dimen1=#2 sp
                  \global\p@intvaluex=\cosine\dimen0
                  \dimen3=\sine\dimen1
                  \global\advance\p@intvaluex by -\dimen3
                  \global\p@intvaluey=\sine\dimen0
                  \dimen3=\cosine\dimen1
                  \global\advance\p@intvaluey by \dimen3
                  }}%
\def\compute@bb{%
                \epsfbbfoundfalse
                \if@bbllx\epsfbbfoundtrue\fi
                \if@bblly\epsfbbfoundtrue\fi
                \if@bburx\epsfbbfoundtrue\fi
                \if@bbury\epsfbbfoundtrue\fi
                \ifepsfbbfound\else\bb@missing\fi
                \ifepsfbbfound\else
                \@latexerr{ERROR. cannot locate BoundingBox}\@whattodobb
                        \@p@@sbbllx{100bp}%
                        \@p@@sbblly{100bp}%
                        \@p@@sbburx{200bp}%
                        \@p@@sbbury{200bp}%
                        \no@bbtrue
                        \psdraft
                \fi
                \count203=\@p@sbburx
                \count204=\@p@sbbury
                \advance\count203 by -\@p@sbbllx
                \advance\count204 by -\@p@sbblly
                \edef\ps@bbw{\number\count203}%
                \edef\ps@bbh{\number\count204}%
                 \edef\@bbw{\number\count203}%
                \edef\@bbh{\number\count204}%
               \if@angle
                        \Sine{\@p@sangle}\Cosine{\@p@sangle}%

{\ps@dimcent=\maxdimen\xdef\r@p@sbbllx{\number\ps@dimcent}%

\xdef\r@p@sbblly{\number\ps@dimcent}%

\xdef\r@p@sbburx{-\number\ps@dimcent}%

\xdef\r@p@sbbury{-\number\ps@dimcent}}%
                        \def\minmaxtest{%
                           \ifnum\number\p@intvaluex<\r@p@sbbllx
                              \xdef\r@p@sbbllx{\number\p@intvaluex}\fi
                           \ifnum\number\p@intvaluex>\r@p@sbburx
                              \xdef\r@p@sbburx{\number\p@intvaluex}\fi
                           \ifnum\number\p@intvaluey<\r@p@sbblly
                              \xdef\r@p@sbblly{\number\p@intvaluey}\fi
                           \ifnum\number\p@intvaluey>\r@p@sbbury
                              \xdef\r@p@sbbury{\number\p@intvaluey}\fi
                           }%
                        \rotate@{\@p@sbbllx}{\@p@sbblly}%
                        \minmaxtest
                        \rotate@{\@p@sbbllx}{\@p@sbbury}%
                        \minmaxtest
                        \rotate@{\@p@sbburx}{\@p@sbblly}%
                        \minmaxtest
                        \rotate@{\@p@sbburx}{\@p@sbbury}%
                        \minmaxtest

\edef\@p@sbbllx{\r@p@sbbllx}\edef\@p@sbblly{\r@p@sbblly}%

\edef\@p@sbburx{\r@p@sbburx}\edef\@p@sbbury{\r@p@sbbury}%
                \fi
                \count203=\@p@sbburx
                \count204=\@p@sbbury
                \advance\count203 by -\@p@sbbllx
                \advance\count204 by -\@p@sbblly
                \edef\@bbw{\number\count203}%
                \edef\@bbh{\number\count204}%
}%
\def\in@hundreds#1#2#3{\count240=#2 \count241=#3
                     \count100=\count240        
                     \divide\count100 by \count241
                     \count101=\count100
                     \multiply\count101 by \count241
                     \advance\count240 by -\count101
                     \multiply\count240 by 10
                     \count101=\count240        
                     \divide\count101 by \count241
                     \count102=\count101
                     \multiply\count102 by \count241
                     \advance\count240 by -\count102
                     \multiply\count240 by 10
                     \count102=\count240        
                     \divide\count102 by \count241
                     \count200=#1\count205=0
                     \count201=\count200
                        \multiply\count201 by \count100
                        \advance\count205 by \count201
                     \count201=\count200
                        \divide\count201 by 10
                        \multiply\count201 by \count101
                        \advance\count205 by \count201
                     \count201=\count200
                        \divide\count201 by 100
                        \multiply\count201 by \count102
                        \advance\count205 by \count201
                     \edef\@result{\number\count205}%
}%
\def\compute@wfromh{%
                \in@hundreds{\@p@sheight}{\@bbw}{\@bbh}%
                \edef\@p@swidth{\@result}%
}%
\def\compute@hfromw{%
                \in@hundreds{\@p@swidth}{\@bbh}{\@bbw}%
                \edef\@p@sheight{\@result}%
}%
\def\compute@handw{%
                \if@height
                        \if@width
                        \else
                                \compute@wfromh
                        \fi
                \else
                        \if@width
                                \compute@hfromw
                        \else
                                \edef\@p@sheight{\@bbh}%
                                \edef\@p@swidth{\@bbw}%
                        \fi
                \fi
}%
\def\compute@resv{%
                \if@rheight \else \edef\@p@srheight{\@p@sheight} \fi
                \if@rwidth \else \edef\@p@srwidth{\@p@swidth} \fi
}%
\def\compute@sizes{%
        \if@scalefirst\if@angle
        \if@width
           \in@hundreds{\@p@swidth}{\@bbw}{\ps@bbw}%
           \edef\@p@swidth{\@result}%
        \fi
        \if@height
           \in@hundreds{\@p@sheight}{\@bbh}{\ps@bbh}%
           \edef\@p@sheight{\@result}%
        \fi
        \fi\fi
        \compute@handw
        \compute@resv
					           \EPS@Width=\@bbw
																\divide\EPS@Width by 1000
   												 \EPS@xscale=\@p@swidth \divide \EPS@xscale by \EPS@Width
					           \EPS@Height=\@bbh
																\divide\EPS@Height by 1000
   												 \EPS@yscale=\@p@sheight \divide \EPS@yscale by\EPS@Height
  \ifnum\EPS@xscale>\EPS@yscale\EPS@xscale=\EPS@yscale\fi
}
\def\psfig{\begingroup\@minisanitize\@@@psfig}
\def\epsfig{\begingroup\@minisanitize\@@@epsfig}

\def\@minisanitize{\@makeother\_\@makeother\:\@makeother\.\@makeother\$}

\def\@@@psfig#1{\vbox {%
        \ps@init@parms
        \parse@ps@parms{#1}%
        \ifnum\@psdraft=1
                \typeout{[\@p@sfilefinal]}%
                \if@verbose
                        \typeout{epsfig: using PSFIG macros}%
                \fi
                \psfig@method
        \else
                \epsfig@draft
        \fi
}
\endgroup
}%

\def\@@@epsfig#1{\vbox {%
        \ps@init@parms
        \parse@ps@parms{#1}%
        \ifnum\@psdraft=1
          \if@angle\use@psfigtrue\fi
          {\ifnum\fig@driver=1\global\use@psfigtrue\fi}%
          {\ifnum\fig@driver=3\global\use@psfigtrue\fi}%
          {\ifnum\fig@driver=4\global\use@psfigtrue\fi}%
          {\ifnum\fig@driver=5\global\use@psfigtrue\fi}%
                \ifuse@psfig
                        \if@verbose
                                \typeout{epsfig: using PSFIG macros}%
                        \fi
                        \psfig@method
                \else
                        \if@verbose
                                \typeout{epsfig: using EPSF macros}%
                        \fi
                        \epsf@method
                \fi
        \else
                \epsfig@draft
        \fi
}
\endgroup
}%

\def\epsf@method{%
        \epsfbbfoundfalse
        \if@bbllx\epsfbbfoundtrue\fi
        \if@bblly\epsfbbfoundtrue\fi
        \if@bburx\epsfbbfoundtrue\fi
        \if@bbury\epsfbbfoundtrue\fi
        \ifepsfbbfound\else\epsfgetbb{\@p@sfile}\fi
        \ifepsfbbfound
           \typeout{<\@p@sfilefinal>}%
           \epsfig@gofer
        \else
          \@latexerr{ERROR - Cannot locate BoundingBox}\@whattodobb
          \@p@@sbbllx{100bp}%
          \@p@@sbblly{100bp}%
          \@p@@sbburx{200bp}%
          \@p@@sbbury{200bp}%
                \count203=\@p@sbburx
                \count204=\@p@sbbury
                \advance\count203 by -\@p@sbbllx
                \advance\count204 by -\@p@sbblly
                \edef\@bbw{\number\count203}%
                \edef\@bbh{\number\count204}%
          \compute@sizes
          \epsfig@@draft
       \fi
}%
\def\psfig@method{%
        \compute@bb
        \ifepsfbbfound
          \compute@sizes
          \psfig@start
          \vbox to \@p@srheight true sp{\hbox to \@p@srwidth true
            sp{\hss}\vss\psfig@end}%
        \else
           \epsfig@draft
        \fi
}%
\def\epsfig@draft{\compute@bb\compute@sizes\epsfig@@draft}%
\def\epsfig@@draft{%
\typeout{<(draft only) \@p@sfilefinal>}%
\if@draftbox
        \hbox{\fbox{\vbox to \@p@srheight true sp{%
        \vss\hbox to \@p@srwidth true sp{ \hss
           {\tt\@p@sfilefinal}
                          \hss }\vss
        }}}%
\else
        \vbox to \@p@srheight true sp{%
        \vss\hbox to \@p@srwidth true sp{\hss}\vss}%
\fi
}%
\psfigdriver{dvips}%
\epsfigRestoreAt

\def\simlt{\stackrel{<}{{}_\sim}}
\def\simgt{\stackrel{>}{{}_\sim}}
\def\ov{\overline}

\def\msbl{m_{\tilde{b}_L}}
\def\mstl{m_{\tilde{t}_L}}
\def\mstr{m_{\tilde{t}_R}}
\def\mtl{m_{\tilde{t}_L}}
\def\mbl{m_{\tilde{b}_L}}
\def\mtr{m_{\tilde{t}_R}}
\def\mbr{m_{\tilde{b}_R}}

\newcommand{\be}{\begin{equation}}
\newcommand{\ee}{\end{equation}}

\newcommand{\bear}{\begin{eqnarray}}
\newcommand{\eear}{\end{eqnarray}}

\def\hu{h_1^{0r}}

\def\IJMPA #1 #2 #3 {Int.~J.~Mod.~Phys.~{\bf A#1}\ (19#2) #3}
\def\MPLA #1 #2 #3 {Mod.~Phys.~Lett.~{\bf A#1}\ (19#2) #3}
\def\NPB #1 #2 #3 {Nucl.~Phys.~{\bf B#1}\ (19#2) #3}
\def\PLB #1 #2 #3 {Phys.~Lett.~{\bf B#1}\ (19#2) #3}
\def\PR #1 #2 #3 {Phys.~Rep.~{\bf#1}\ (19#2) #3}
\def\PRD #1 #2 #3 {Phys.~Rev.~{\bf D#1}\ (19#2) #3}
\def\PTP #1 #2 #3 {Prog.~Theor.~Phys.~{\bf #1}\ (19#2) #3}
\def\PRL #1 #2 #3 {Phys.~Rev.~Lett.~{\bf#1}\ (19#2) #3}
\def\RMP #1 #2 #3 {Rev.~Mod.~Phys.~{\bf#1}\ (19#2) #3}
\def\ZPC #1 #2 #3 {Z.~Phys.~{\bf C#1}\ (19#2) #3}

\begin{document}

\begin{titlepage}

\title{\bf The baryogenesis window in the MSSM}

\author{{\bf B. de Carlos} \\
Centre for Theoretical Physics, University of Sussex\\
Falmer, Brighton BN1 9QH, UK \\
and \\
{\bf J.R. Espinosa} \\ 
Department of Physics and Astronomy, University of Pennsylvania\\
Philadelphia PA 19104-6396 USA}

\date{} 
\maketitle
\vspace{.5cm}
\def\baselinestretch{1.15}
\begin{abstract}
Thermal two-loop QCD corrections associated with light stops have a 
dramatic effect on the strength of the MSSM electroweak phase 
transition, making it more strongly first order as required for the 
viability of electroweak baryogenesis. We perform a perturbative 
analysis of the transition strength in this model, including these 
important contributions, extending previous work to arbitrary values 
of the pseudoscalar Higgs boson mass, $m_A$. We find a strong enough 
transition in a region with $2\simlt \tan\beta\simlt 4$ and $m_A\simgt 
120\ GeV$, a light Higgs boson with nearly standard couplings, and 
mass below $85\ GeV$ within the reach of LEP~II, and one stop not 
much heavier than the top quark. In addition, we give a qualitative 
discussion of the parameter space dependence of the transition 
strength and comment on the possibility that the transition turns to 
a crossover for sufficiently large Higgs masses.
\end{abstract} 
\vspace{4cm}
\leftline{March 1997}

\thispagestyle{empty}

\vskip-23.cm
\rightline{{ SUSX--TH--97--005}}
\rightline{{ UPR--0740-T}}
\rightline{{ IEM--FT--149/97}}
\vskip3in

\end{titlepage}

\def\baselinestretch{1.1}
\section{Introduction}
It is by now well established that successful electroweak 
baryogenesis \cite{ewb} calls for new physics at the Fermi scale. 
In the minimal Standard Model (SM) it is not possible 
to generate a sufficient baryon asymmetry at the electroweak phase 
transition (EWPHT) because the necessary CP violation is far too 
small. In addition, the phase 
transition is at best weakly first order for realistic values of the 
Higgs mass, and any net $B+L$ number created would be subsequently 
erased by unsuppressed sphaleron 
processes in the broken phase. This failure could be an indication that 
some type of $B-L$ violating new physics is at work above the 
electroweak scale (see e.g.~\cite{bu}). Although that is an 
interesting possibility, we are not really forced to give up the 
beautiful idea of electroweak baryogenesis, because physics at the 
$100\ GeV$ scale may not be described by the SM. Supersymmetry (SUSY)  
\cite{susy}, the most promising candidate for physics beyond the SM, 
is expected to be relevant at such energies. 
Thus, it is natural to examine the prospects for electroweak 
baryogenesis in models of low-energy SUSY. This is particularly 
interesting for two reasons; first, because low-energy SUSY is well 
motivated for reasons not related to the matter asymmetry problem. 
Second, the parameter regions where electroweak baryogenesis 
may be successful are in the reach of the present generation
of colliders so that the viability of the scenario would be tested in 
a not too distant future.

Supersymmetric electroweak baryogenesis has been the subject of many 
studies in recent years. We restrict ourselves to  the simplest   
realization of low-energy SUSY, the Minimal Supersymmetric Standard Model (MSSM). 
Non-minimal models\footnote{In models without $R$--parity conservation, 
baryogenesis before the EWPHT faces the problem of baryon depletion by
the combined action of sphalerons and $B-L$ violating processes. 
Electroweak baryogenesis offers a particularly appealing solution to 
this problem, although other possible solutions have been proposed 
\cite{dahe}.} have also been considered \cite{n-mssm}.

In SUSY models the presence of extra 
sources of CP violation \cite{cp} besides the Kobayashi-Maskawa angle
in the SM can be enough for electroweak baryogenesis and imply CP 
violating signals on the verge of being seen in 
upcoming experiments. The observed baryon asymmetry can be accounted for 
if the new CP violating phases are greater than $10^{-(2-4)}$ and some 
superpartners are light enough\footnote{These results are affected if 
the rate of baryon number violation in the high temperature symmetric
phase is smaller than it was previously thought to be (see 
\cite{a5T4,a5t4m,a5t4n}).} \cite{hune} (see also \cite{bgmssm,cqrvw,wo}). 
In addition, the presence of two Higgs 
doublets makes possible the spontaneous violation of CP in the Higgs
sector  at finite temperature \cite{copi} (spontaneous breaking 
does not occur at $T=0$ for realistic values of the parameters of the 
model \cite{po}). This mechanism takes place for small values of the 
pseudoscalar Higgs boson mass $m_A$ and large values of $\tan\beta$ 
(the ratio of Higgs vevs), and could play an important r$\hat{o}$le 
for electroweak baryogenesis.

The requirement of a sufficiently strong first-order transition to avoid
$B+L$ sphaleron erasure of the asymmetry sets a strong upper limit on the
mass of some Higgs boson in the theory. Therefore, it is crucial to 
determine the strength of the electroweak transition and identify the 
regions of parameter space where it can be strong enough. More precisely, 
the requirement is
\cite{shap}
\be
\label{vt}
\frac{v(T_c)}{T_c}>1,
\ee
where $T_c$ is the critical temperature of the transition, defined by 
the coexistence of two degenerate minima, and $v=\sqrt{v_1^2+v_2^2}$, the order 
parameter of the transition, is the vacuum expectation value (vev) of the Higgs 
field driving electroweak symmetry breaking (normalized to $v=246\ GeV$ at $T=0$). 
The first studies of this problem in the MSSM have already pointed out that 
(\ref{vt}) would require light stops in the spectrum (that is, with masses 
comparable to the critical temperature of the transition). In ref.~\cite{gi}, 
it seemed inplausible that these constraints were satisfied, so MSSM electroweak
baryogenesis was considered unlikely. On the other hand, 
ref.~\cite{my} found a sizeable region of parameter space with light 
stops and large mass $m_A$ for the pseudoscalar Higgs boson, in which 
the transition was strong enough for baryogenesis. However, a careful 
treatment of the LEP bounds on the Higgs mass for large $m_A$ reduces 
that region significantly. Both \cite{gi} and \cite{my} 
studied the one-loop effective potential perturbatively at finite 
temperature without resummation of higher loop contributions.

A more careful analysis of this potential, including resummation of
the so--called Daisy diagrams which have an important effect, was 
carried out in \cite{eqz} (for large $m_A$) and \cite{beqz} (for 
arbitrary $m_A$). A small region of parameter space in which the transition
is strong enough remained after imposing experimental constraints.
This region corresponded to large $m_A$ (which in turn determines a Higgs 
spectrum with only a light Higgs boson), small $\tan\beta$ 
($\tan\beta\sim 2$), light stops compatible with 
experimental limits and with negligible L--R mixing, heavy gluinos, and
the light Higgs barely above the LEP lower limit ($m_h>65\ GeV$ at that time). 
The rest of the supersymmetric particles do not influence much the 
transition so that their spectrum is not constrained, although the 
general tendency is to prefer heavy superpartners. Even if this is a
significant improvement with respect to the SM, the window for 
baryogenesis was clearly small and is already excluded after the improvement in
the Higgs mass bound from LEP ($m_h>70.7\ GeV$ from ALEPH \cite{ALEPH}).

The subject was revived recently by refs.~\cite{cqw,esp}. In 
ref.~\cite{cqw}, the region of light masses for $\tilde{t}_R$ was
increased by taking negative values of its soft-mass squared,
$m_U^2<0$. Negative values of $m_U^2$ are associated with the existence 
of dangerous colour-breaking minima in the MSSM multi-scalar potential. 
While in \cite{eqz,beqz} the lower limit on $m_U^2$ was set to zero, 
\cite{cqw} shows that moderately negative values of $m_U^2$ can still
be cosmologically allowed; the electroweak minimum is the deepest one, 
and is the one populated when the Universe cools down. By moving into 
a region with smaller $m_{\tilde{t}_R}$ (while keeping negligible 
mixing), the electroweak transition is stronger and sphaleron erasure 
can be avoided (see also \cite{dege}), thus enlarging the region where 
electroweak baryogenesis could take place.

In ref.~\cite{esp}, it was shown that two-loop QCD corrections 
associated with light stops dramatically affect the quantitative 
behaviour of the effective potential. Inclusion of these contributions
is then mandatory for a more precise study of the transition, and these
corrections make it strong enough for baryogenesis even if $m_U^2>0$. 
Therefore, this effect improves the situation in all the regions of 
parameter space with light stops. 

Both \cite{cqw} and \cite{esp} focused on the large $m_A$--small 
$\tan\beta$ region of parameter space preferred for baryogenesis and 
used perturbative calculations of the temperature dependent effective 
potential. As discussed in \cite{cqw,esp,espcos}, the perturbative treatment 
is expected to give reliable results for Higgs masses larger than in the 
SM case (where the limit for the validity of perturbative calculations
was roughly around $m_W$). A very useful complementary approach was 
followed in refs.~\cite{laine,falo,clka}, where 3d effective theories 
\cite{3d1,3d2} were constructed to study the high temperature phase of
the MSSM and, in particular, the nature of the electroweak transition.
For not too small $m_U^2$, the relevant 3d effective theory is well 
approximated by an $SU(2)$ $+$ Higgs model (as in the SM but with 
different couplings) for which there exist non--perturbative lattice 
studies. The results of these analyses are in reasonable 
agreement\footnote{ Some of the results obtained in \cite{clka} were 
artifacts of an implicit expansion in the stop mixing over the Debye 
masses and disappear when the convergence of this expansion is under 
control \cite{cline}. In particular, the baryogenesis region of small 
$m_A$--large $\tan\beta$ found in \cite{clka} does not subsist.} with 
those of \cite{cqw,esp}. A region with a strong transition is 
confirmed for light stops (in particular light $\tilde{t}_R$), small 
stop mixing, small $\tan \beta$, and Higgs masses up to about 
$80\ GeV$.

In any case, it is important to keep in mind that for small values of 
$m_U$ (the interesting region for baryogenesis) the relevant 3d effective 
theory is no longer $SU(2)$ + Higgs, but $\tilde{t}_R$ and gluons should 
also be taken into account. The construction of such an effective theory is 
straightforward and was sketched in \cite{laine}, but Monte Carlo 
simulations for that theory are not available. 
Another possible shortcoming of the existent 3d reduction 
studies may show up in the small $m_A$ region where the two Higgs 
doublets are light. In such a case the effective theory may require 
the presence of both Higgses; it may not be justifiable to 
integrate out the heavier of the two. This point is discussed in more 
detail in the next sections. 

The purpose of this paper is to further continue the perturbative 
exploration of the MSSM baryogenesis window in the region of small and
moderate $m_A$, including the important two-loop corrections. As we will 
show, smaller $m_A$ always corresponds to a weakening
of the transition so that (for fixed values of other parameters) there
is a lower bound on $m_A$ below which the transition is too weak for 
baryogenesis. This lower limit decreases for smaller $\tan\beta$.

The small $m_A$ case is interesting for several reasons. First, it can
be preferred in some mechanism for the generation of baryon number 
\cite{hune}. Second, in the case of small $m_A$--large 
$\tan\beta$, the spontaneous CP violation mechanism of 
ref.~\cite{copi} already mentioned could play an important 
r$\hat{o}$le in baryogenesis. Unfortunately, we do not find any 
allowed region for large values of $\tan\beta$. Finally, 
the 3d approach in this region may be less straightforward than for 
large $m_A$. In particular, contact with the Monte Carlo lattice 
results for the $SU(2) +$ Higgs theory could be hampered by the 
necessity of keeping both Higgses in the effective theory.

In this paper we are going to make a 
purely perturbative analysis of the transition, so we take some time to 
discuss in section~2 the range of parameter space where 
non-perturbative effects start to play a non-negligible r$\hat{o}$le 
in the phase transition. In analogy with the SM case, we expect that 
this range is associated with the change in the nature of the 
EWPHT from first order to a crossover. We discuss in section~3 the 
expectations for a strong phase transition in the ($m_A,\tan\beta$) 
parameter space making a simple qualitative analysis of the effective
potential,
both at zero temperature and including the dominant temperature 
corrections. In section~4, we write the high temperature effective 
potential at one-loop order with Daisy resummation, concentrating on 
the small $m_A$ region. In section~5, we consider the effect of dominant 
contributions to the two-loop resummed potential. We present our results 
in section~6 and our conclusions in section~7. 
In appendix~A, we collect field dependent 
masses and mixing angles (both at zero and finite T) for the particles
relevant to our study. Appendix~B gives the lengthy expression for 
the non-expanded full two-loop potential. Finally, appendix~C contains
the dominant contributions to the two-loop resummed potential in a 
high T expansion. 

\vspace{0.5cm}
\section{Crossover in the MSSM}

In the MSSM with 
light stops, the expansion parameter of the resummed perturbation 
theory at finite temperature is $\epsilon_{MSSM}\sim h_t^2/\lambda$ 
rather than $\epsilon_{SM}\sim g^2/\lambda$ \cite{cqw,esp,espcos}, 
where $h_t$ is the top 
Yukawa coupling, $g$ the $SU(2)_L$ gauge coupling and $\lambda$ the 
quartic Higgs boson coupling. Perturbation 
theory is expected to be applicable to the study of the EWPHT in the 
MSSM for a range of Higgs masses wider than in the SM, with a 
critical upper limit governed by $m_t$ instead of $m_W$. We present  
a rough estimate of a related quantity, the critical Higgs mass 
beyond which the phase transition changes from first--order to a 
crossover. In the SM, it is possible to estimate this mass analytically 
by imposing the equality of the
transverse $W$ mass in the broken and 
symmetric phases at the critical temperature \cite{cresti}. The argument 
arises naturally in the context of the 3d reduced theory but we use
4d quantities in what follows. The mass in the broken phase is given by
\be
\label{mwbr}
{m_W(T_c)}_{br}=\frac{1}{2}gv(T_c),
\ee
where the vev $v(T_c)$ can be obtained from the effective potential
\be
\label{poten}
V \simeq \frac{1}{2}m^2(T)\varphi^2 + \frac{1}{8}\lambda(T)\varphi^4
-\frac{T}{16\pi} g^3 \varphi^3.
\ee
The mass in the symmetric phase, the magnetic mass, is of 
non-perturbative origin and can be estimated by solving a set of gap 
equations \cite{gap1,gap2}. It is
\be
\label{mwsym}
{m_W(T_c)}_{sym}=C g^2 T_c,
\ee
where $C=0.28$ in the Standard Model \cite{gap2} (it is basically equal to its 
value in the $SU(2)$--$\sigma$ model). Equating (\ref{mwbr}) and 
(\ref{mwsym}) and solving for $\lambda$ one arrives at the critical 
mass for the onset of crossover
\be
m_h^c=\sqrt{\frac{3}{4\pi C}} m_W\sim 74\  GeV \; .
\ee
This naive estimate is close to the numerical result of
lattice simulations \cite{cross} which give $m_h^c\sim 80\ GeV$.

In the MSSM case with large $m_A$ and light 
stops, we add to the potential (\ref{poten}) the stop contribution,
roughly approximated by
\be
\delta V \simeq - r \frac{T}{4\pi\sqrt{2}}h_{t,SM}^3\varphi^3,
\ee
where $h_{t,SM}=h_t\sin\beta$, and the constant $r$ parameterizes the 
effective strength of the stop corrections. It is normalized in such
a way that it would be equal to 1 if all screening effects 
from soft and thermal masses were negligible. In realistic cases $r\ll 1$. 
Heavy soft masses or light gluinos (which give a sizeable 
contribution to the stop thermal mass) would decrease $r$. On the other
hand, a negative value for $m_U^2$ tends to increase $r$. From 
the numerical study of the transition, we can estimate that $r\sim 0.2$ 
when stops affect the transition sizeably without conflicting with experimental
constraints.
\begin{figure}[hbt]
\centerline{
\psfig{figure=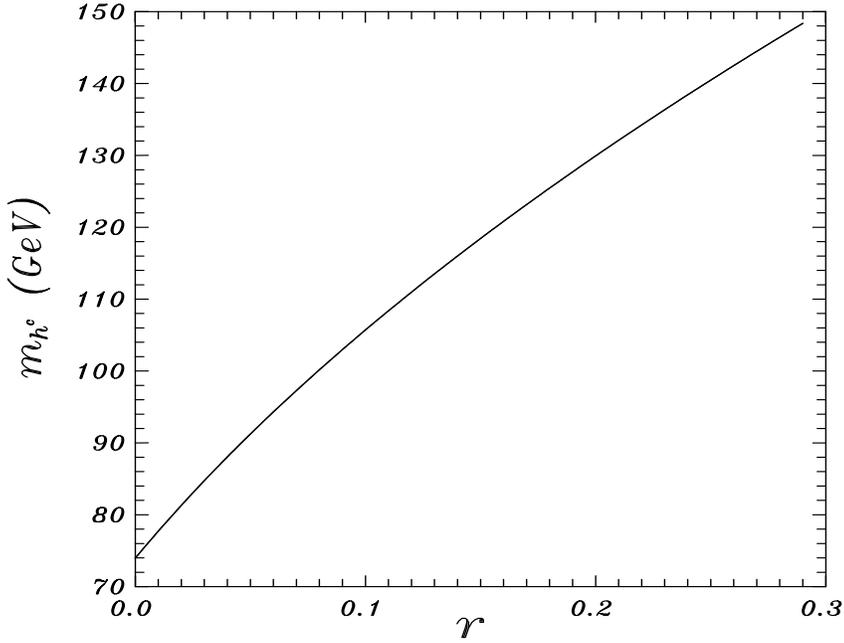,height=10cm,width=12cm,bbllx=2cm,bblly=2cm,bburx=15cm,bbury=15cm}}
\caption{\footnotesize
Critical Higgs boson mass for crossover in the MSSM [as estimated by 
formula (\ref{mhcmssm})]. The parameter $r$ measures the effect of
stops on the transition: $r=0$ corresponds to the pure SM limit
and realistic values of $r$ should not be larger than $0.2-0.3$.} 
\end{figure} 

The magnetic mass does not depend on the presence of stops (in the 
same way that it is not sensitive to the Higgs mass in the SM) and 
it should be well approximated by the same expression (\ref{mwsym}). 
The following estimate for $m_h^c$ in the MSSM follows
\be
\label{mhcmssm}
m_h^c=m_W\sqrt{\frac{3}{4\pi C}\left(
1+r\frac{m_t^3}{m_W^3}
\right)}.
\ee
Eq.~(\ref{mhcmssm}) is plotted as a function of the parameter $r$ 
in figure~1. Due to the $m_t^3$ factor, $m_h^c$ grows very rapidly
with $r$; for example, $r\sim 0.1$ corresponds to $m_h^c\sim 100\ GeV$. 
Below the curve the electroweak phase transition is first-order, 
and weakens as the curve is approached. Above the line there is not a 
phase transition, but an analytic crossover.

To interpret figure~1 correctly one needs to take into account that a 
change in the masses of the stops affects $r$ directly, and $m_{h}$ 
through radiative corrections. Moreover, we have obtained 
eq.~(\ref{mhcmssm}) for the case of large $m_A$, when $\lambda$ in 
(\ref{poten}) can be directly related to the light Higgs boson
mass. For lower $m_A$ one can still obtain a critical value $\lambda^c$ for 
$\lambda$, but its relation to $m_h^c$ involves the 
mixing angle between the two Higgs scalars (and $m_h^c\leq 
\lambda^cv$). Also, the important effect of two-loop corrections on the strength
of the transition cannot be accounted for by a simple redefinition of $r$.
The results presented in figure~1 are thus a rough estimate of the critical mass
values expected.
 An alternative way to describe the emergence of 
crossover is to equate (\ref{mwbr}) and 
(\ref{mwsym}) to obtain $v/T_c\sim 2Cg\sim 0.4$, and interpret this 
as the critical value for the jump in the order parameter as computed 
in perturbation theory. When perturbatively the transition is so weak, 
non-perturbative effects can no longer be neglected and they would 
eventually change the transition from first order to a crossover.
This definition can be used now with the two-loop corrected $v/T_c$.
As we are primarily motivated by the possibility of electroweak baryogenesis,
which requires a strong transition, we do not attempt here a more 
precise determination of $m_h^c$ along these lines.

\vspace{0.5cm}
\section{Parameter Space}

Light stops increase the strength of the electroweak phase transition 
in the MSSM. Whether this effect is large enough to permit 
baryogenesis depends on the details of the Higgs 
sector at $T=0$. At tree-level, this sector is completely determined 
by two parameters: the pseudoscalar mass $m_A$, and $\tan\beta$, the 
ratio of Higgs vevs at zero $T$. In this section we explore the plane 
$(m_A,\tan\beta)$ showing the expectations for a strong phase 
transition.

\vspace{0.3cm}

$\em 1)$ Large $m_A$ $(m_A\gg m_Z,T_c)$.

\vspace{.2cm}

In this case only one Higgs doublet remains at the electroweak scale 
and is related to symmetry breaking, while the other is heavy and 
decouples from the problem. The Higgs potential relevant to the study 
of the electroweak transition is SM-like, $V(\varphi)$, with 
$\varphi=\varphi_1\cos\beta+\varphi_2\sin\beta$, but the light Higgs 
mass is not a free parameter as in the SM and it is determined  
at one--loop \cite{radcor} to be
\be
\label{mh}
m_h^2=m_Z^2\cos^22\beta +
\frac{3}{2\pi^2}\frac{m_t^4}{v^2}\log\frac{m_{\tilde{t}_1}
m_{\tilde{t}_2}}{m_t^2} + f(A_t+\mu/\tan\beta).
\ee
The logarithmic radiative correction increases with increasing stop 
masses and can be sizeable. The last term in (\ref{mh}) depends on 
the stop mixing through the combination $X_t=A_t+\mu/\tan\beta$ and 
reads:
\be
\label{mixcor}
f(X_t)=\frac{3}{4\pi^2}\frac{m_t^4}{v^2}X_t^2\left[
2h(m_{\tilde{t}_1}^2,m_{\tilde{t}_2}^2)
+X_t^2f(m_{\tilde{t}_1}^2,m_{\tilde{t}_2}^2)
\right],
\ee
where
\be
h(a,b)=\frac{1}{a-b}\log\frac{a}{b},
\ee
and
\be
g(a,b)=\frac{1}{(a-b)^2}\left[2-\frac{a+b}{a-b}\log\frac{a}{b}\right].
\ee
The correction (\ref{mixcor}) is zero if the stop mixing is negligible 
("zero mixing" case), reaches a positive maximum for large values of the 
mixing ("maximal mixing" case), then drops (even getting negative) 
if the mixing is further increased ("extreme mixing" case).

As in the SM case, the phase transition becomes weaker for larger $m_h$,
so the best region for baryogenesis always corresponds to a 
region with lightest possible Higgs mass\footnote{The
fact that the light Higgs
is SM--like implies that the usual LEP bound on its mass ($m_h>70.7\ 
GeV$) also applies; for large $m_A$ there is no possibility of 
accessing lower values of $m_h$ to strengthen the 
transition.}. From (\ref{mh}), we see that the region preferred to get
a strong transition is one with small $\tan\beta$, and light stops with 
negligible mixing (the case of a light Higgs due to extreme mixing 
will be discussed later). Of course, it is not enough to have a light 
Higgs if no effect beyond the SM helps to strengthen the transition. 
There are two benefits of having light stops; first, the radiative 
corrections to the Higgs mass are smaller for lighter stops so that 
the Higgs boson can be lighter. Second, direct stop contributions 
to the $T\neq 0$ potential enhance the transition strength (for small
mixing). These ingredients describe the well known region preferred for 
baryogenesis. We now examine other possible effects to see 
if we can find additional regions supporting a strong phase
transition.

\vspace{.3cm}

$\bullet$ {\em Intermediate mass stops}

If the stops are very heavy compared to the transition temperature, 
their thermal contributions are Boltzmann suppressed and the SM case
for the corresponding $m_h$ is recovered. On the other hand, moderately
heavy stops can still influence the transition via a $T=0$ effect 
described in \cite{anha} (see also the discussion in \cite{gi}). If we 
neglect all couplings other than $h_t$ in the loop corrections, the 
one-loop $T=0$ potential can be written approximately as 
\bear
\label{vanha}
V(\varphi)&=&-\frac{1}{2}m^2\varphi^2+\frac{1}{4}\lambda\varphi^4\nonumber\\
&+&\frac{6}{64\pi^2}\left[
m_{\tilde{t}_1}^4\left(\log\frac{m_{\tilde{t}_1}^2}{Q^2}-\frac{3}{2}\right)
+m_{\tilde{t}_2}^4\left(\log\frac{m_{\tilde{t}_2}^2}{Q^2}-\frac{3}{2}\right)
-2m_t^4\left(\log\frac{m_t^2}{Q^2}-\frac{3}{2}\right)
\right],
\eear
where $m_{\tilde{t}_{1,2}}$ and $m_t$ are field-dependent quantities
and can be found in appendix~A (making the
replacement $\varphi_1\rightarrow \varphi\cos\beta$, 
$\varphi_2\rightarrow\varphi\sin\beta$, appropriate in the large $m_A$
limit). When stops are heavy (for example, 
with equal soft masses $m_Q=m_U\equiv M_S$) compared to $m_t$, it is
possible to expand their contribution to the potential in powers of
$\varphi/M_S$. If we express 
$m^2$ and $\lambda$ of eq.~(\ref{vanha}) in terms of  
the radiatively corrected vev $v$ and mass $m_h$, the potential takes the
simple 
form
\be
V(\varphi)=\frac{1}{4}\varphi^2(\varphi^2-v^2)\frac{m_h^2}{v^2} +\Delta 
V_t(\varphi) + \Delta V_{\tilde{t}}(\varphi),
\ee
with
\be
\Delta V_t(\varphi)=-\frac{m_t^4}{\pi^2}\frac{\varphi^2}{v^2}\left[
1+\frac{\varphi^2}{v^2}\left(
\log\frac{\varphi^2}{v^2}-\frac{3}{2}
\right)
\right],
\ee
and
\be
\Delta 
V_{\tilde{t}}(\varphi)=\kappa\frac{m_t^4}{\pi^2} 
\frac{\varphi^2}{M_S^2}\left(3-6\frac{\varphi^2}{v^4}
+4\frac{\varphi^4}{v^4}\right) + {\cal O}(\varphi^8/M_S^4).
\ee
Here 
\be
\label{kappa}
\kappa=1-\frac{3X_t^2}{2M_S^2}.
\ee
$\Delta V_{t,\tilde{t}}(\varphi)$ satisfy $\Delta V'=\Delta V''=0$ at 
$\varphi=v$ so that they do not affect $v$ or $m_h$. Although $\Delta 
V_{\tilde{t}}$ shows the decoupling explicitly, it represents a 
deviation from the pure SM potential for not so large values of
$M_S$. This 
effect corrects the transition strength for a fixed Higgs mass. For 
example, if $M_S=500\ GeV$ the increase in $v(T_c)/T_c$ with respect 
to the SM with the same Higgs mass can be estimated to be a factor 
$1+0.1\kappa$ larger. This is a $10\%$ effect if the stop mixing is 
negligible. Unfortunately, this modest increase for heavy stops is 
counterbalanced by the direct increase of $m_h$ through radiative 
corrections so there is no net gain, and larger stop masses always
give a weaker transition. On the other hand, it is apparent from 
(\ref{kappa}) that stop mixing tends to weaken the enhancement of 
$v(T_c)/T_c$, eventually having a negative effect\footnote{The rest of squarks and sleptons give only a small 
correction (see \cite{gi}) through this effect. On the other hand, if they 
are lighter and in thermal equilibrium they raise the screening masses of 
stops and thus have a negative effect on the transition strength (see 
however \cite{falo}).}.

\vspace{0.3cm}

$\bullet$ {\em Extreme stop mixing}

In this case, one may hope to get small masses for the Higgs boson 
via a negative radiative correction from the last term in
(\ref{mh}). To reach this region either one or both stops should 
be heavy, otherwise the large mixing would drive the lighter stop mass
to an imaginary value. When both stops are heavy, the situation is
described by 
the preceding paragraph. If one stop remains light (say $\tilde{t}_R$,
to avoid problems with $\Delta\rho$), its mass squared takes the form 
\be
\label{extr}
m^2_{\tilde{t}_1}=m_U^2+\frac{1}{2}h^2_t\sin^2\beta\varphi^2\left(
1-\frac{X_t^2}{m_Q^2}\right)+D_R(\varphi),
\ee
in which $D_R(\varphi)$ is the $D$ term contribution, and $m_U$ ($m_Q$) is
the $\tilde{t}_R$ ($\tilde{t}_L$) soft mass. Eq.~(\ref{extr}) shows
that the coupling to the Higgs field $\varphi$ is reduced by the 
mixing, and the influence of this stop on the transition diminishes 
\cite{cqw}. Thus, no improvement is expected over the SM in this region.

\vspace{0.3cm}

$\bullet$ {\em Stop mixing two-loop effects}

These corrections, calculated in \cite{esp}, can have two different 
effects. First, the stop mixing angle $\alpha_t$ appears in the stop 
contributions. This changes the corrections slightly but no dramatic 
enhancement appears. The second effect, potentially more important, 
arises from a set of corrections that depend on the new trilinear 
coupling $h-\tilde{t}-\tilde{t}$. The dominant contribution of setting
sun diagrams involving this coupling is \cite{esp}: 
\bear
\delta V&=&
\frac{N_cT^2}{16\pi^2}(h_t\sin\beta X_t)^2\left[
\log\frac{\ov{m}_h+\ov{m}_{\tilde{t}_1}+\ov{m}_{\tilde{t}_2}}{3T}+
\log\frac{\ov{m}_G+\ov{m}_{\tilde{t}_1}+\ov{m}_{\tilde{t}_2}}{3T}\right.
\nonumber\\
\label{twomix}
&+&\left.  \log\frac{\ov{m}_G+\ov{m}_{\tilde{t}_1}+\ov{m}_{\tilde{b}_L}}{3T}
+\log\frac{\ov{m}_G+\ov{m}_{\tilde{t}_1}+\ov{m}_{\tilde{b}_L}}{3T}\right].
\eear
Barred scalar masses in this expression include effects from thermal
screening. We 
have omitted in (\ref{twomix}) other contributions proportional to 
gauge couplings or $h_b$, as well as terms of the type 
$\log[(m_h+2m_{\tilde{t}_1})/(m_h+2m_{\tilde{t}_2})]$, that have a
small effect on $v/T_c$ (a more refined approximation can be obtained 
from ref.~\cite{esp}).

It is a simple exercise to show that a contribution to the potential
of the form $\delta V=K\log[(\varphi^2+\Pi)/\Pi]$ increases $v/T_c$ 
for positive $K$ (see \cite{arnes}). The contribution (\ref{twomix}), 
which can be approximated by an expression of this form, tends then 
to raise $v/T_c$. However, keeping in 
mind that $X_t$ cannot be made much larger than the scale $T$, rough
numerical estimates show that this two-loop enhancement of $v/T_c$ is
below the few percent level, while the negative effect of a non zero
$X_t$ at one-loop level \cite{eqz} is quite significant.

\vspace{0.3cm}

$\em 2)$ Small $m_A$ $(m_A\sim m_Z, T_c)$.

\vspace{.2cm}

For low $m_A$ the transition is not forced to proceed along the 
fixed direction $\varphi_2/\varphi_1=\tan\beta$. Consider the 
($T\neq 0$) mass matrix at the origin $\varphi_1=\varphi_2=0$ and 
define the field-direction of the lowest eigenvalue as 
$\varphi=\varphi_1\cos\theta(T)+\varphi_2\sin\theta(T)$. The behaviour
of $\theta(T)$ has been studied in refs.~\cite{beqz,falo}. At $T=0$ it
is given by 
\be
\label{delt}
\tan 2\theta \simeq \frac{m_A^2\tan 2 
\beta}{m_A^2+M_Z^2+\Delta m^2/\cos 2\beta},
\ee
where $\Delta m^2>0$ is a one-loop correction term that can be found 
in \cite{beqz}. We see that $\tan \theta\geq \tan\beta$ and for 
sufficiently low $m_A$, $\tan \theta\gg \tan\beta$. When $T$ increases,
the radiative corrections in (\ref{delt}) receive a negative 
contribution that grows like $h_t^2T^2$, which decreases 
$\tan\theta(T)$. The critical temperature for the transition 
is usually reached for $\tan \theta(T_c)\gg \tan\beta$ (see 
\cite{beqz}).\footnote{In the language of 3d effective theories, the 
heavy Higgs field that can be integrated out at the moment of the 
transition is basically $H_1$ for small $m_A$.} 

In the low $m_A$--low $\tan\beta$ ($\sim 1$) region, $\theta(T_c)$ 
could be below $\tan\beta$ \cite{falo}. In general, $\tan\theta(T_c)$ is 
approximately equal to $\varphi_2(T_c)/\varphi_1(T_c)$ \cite{beqz} 
with the larger differences being expected for low $\tan\beta$. 
If $\tan\theta(T_c)$ and $\varphi_2(T_c)/\varphi_1(T_c)$ differ
significantly, the usual procedure of integrating out the heavy Higgs at
$T_c$ may not be justified in the construction of 3d effective theories, 
at least to study $v/T_c$. However, as already noted in \cite{falo}, this
discrepancy is most significant for values of $m_A$ and $\tan\beta$
excluded experimentally. Nevertheless, it is important to keep this
possible complication in mind when constructing 3d effective theories for
models with a non--minimal Higgs sector.

In the most common case, $\theta(T_c)\sim
\varphi_2(T_c)/\varphi_1(T_c)$, and we can safely assume that the 
transition takes place along that direction. In particular, we note 
that the Higgs quartic coupling relevant to determine the strength of 
the transition is the quartic self coupling of that particular 
direction. In other words, the Higgs mass whose magnitude controls the
transition strength is some effective mass for excitations along the 
field direction excited in the transition\footnote{A similar 
observation is helpful to estimate rather accurately the sphaleron 
mass in the MSSM \cite{moq}.}. In principle, this is not the mass of 
any of the physical Higgses. It is straightforward to obtain an expression
for this effective mass which is the one written in 
eq.~(\ref{mh}) with the replacement $\beta\rightarrow\theta(T_c)$. At 
low $m_A$, the physical mass of the lightest Higgs is below (\ref{mh}) 
and the corresponding eigenstate is associated with excitations in a field 
direction different than $\theta(T_c)$.

If one starts with some $v/T_c$ for large $m_A$ and fixed $\tan\beta$, the
two effects just described cooperate to decrease $v/T_c$ as $m_A$ lowers; 
lower values of $m_A$ increase $\tan\theta(T_c)$, which probes larger 
effective masses and thus weakens the transition. The maximal value of 
(\ref{mh}) is already saturated for large $\tan\beta$,
so $\tan\theta(T_c)\gg \tan\beta$ does not make much difference in that
case: in the large $\tan\beta$ regime the transition tends to have $v/T_c$
small and independent of $\tan\beta$. This feature is actually observed in
refs.~\cite{laine,falo}. The only low $m_A$ region where one can hope for
a strong transition is once again the region with low $\tan\beta$,
although it is expected to be reduced with respect to the large $m_A$ case. 

The qualitative behaviour described in the previous discussions is 
confirmed by numerical computations of $v/T_c$. In the rest of
the paper, we focus on the small $m_A$ region and compute $v/T_c$ 
by studying the effective potential up to two loop order in 
resummed perturbation theory.

\vspace{.5cm}
\section{One-loop resummed potential} 

In this paper, we assume that the only particles present at the
energy scale of the electroweak phase transition are the SM particles 
plus the extra MSSM Higgs doublet and squarks of the third generation 
[$\tilde{Q}_L=(\tilde{t}_L,\tilde{b}_L)$ and $\tilde{t}_R,\tilde{b}_R$].
The masses of the rest of the supersymmetric particles are assumed to be
large compared to the critical temperature of the transition and their 
contributions are Boltzmann-suppressed. This is just a simplifying 
assumption and our results are qualitatively the same for more general 
spectra. The most important point is that gluinos should be heavy and 
Boltzmann-decoupled. If not, they would give significant contributions 
to the thermal masses for stops \cite{eqz,beqz}, making 
them heavier and weakening the phase transition. The 
details of the transition depend weakly on the rest of the spectrum.  
Our model gives a fair estimate of the phase transition strength for 
realistic supersymmetric spectra (in particular models with a neutral 
LSP), with the only condition that gluinos should be heavy.

For general values of $m_A$, the effective potential is a function of two
Higgs background fields: $\varphi_1= \langle H_1^0\rangle \sqrt{2}$ and 
$\varphi_2= \langle H_2^0\rangle \sqrt{2}$, in which $H_i^0$ are the
neutral components of the two MSSM Higgs doublets. At $T=0$, 
$\varphi_2/\varphi_1=v_2/v_1=\tan \beta$ and $v_1^2+v_2^2 = 
(246\ GeV)^2$. 

The tree--level effective potential is
\be
V_0(\varphi_1,\varphi_2) = \frac{1}{2} m_1^2 \varphi_1^2 + \frac{1}{2}
m_2^2 \varphi_2^2 + m_{12}^2 \varphi_1 \varphi_2 + \frac{g^2+g'^2}{32}
(\varphi_1^2-\varphi_2^2)^2 \;\;,
\ee
with the tree-level quartic couplings fixed by SUSY in terms of gauge 
couplings as shown. 

At one--loop (with resummation of Daisy diagrams), the effective 
potential (written in $\overline{\rm MS}$ scheme and 't Hooft-Landau 
gauge) receives the contribution: 
\be
V_1(\varphi_1,\varphi_2)  =  \frac{1}{2 \pi^2} \sum_i n_i \left\{ 
\frac{1}{32} m_i^4(\varphi_1,\varphi_2) \left[ \log 
\frac{m_i^2(\varphi_1,\varphi_2)}{Q^2} - C_i \right] 
 +  T^4 \widetilde{J}_i \left[ \frac{m_i^2(\varphi_1,\varphi_2)}{T^2} 
\right] \right\} \;\;.
\ee
The field--dependent masses $m_i^2(\varphi_1,\varphi_2)$ for gauge 
bosons, top and bottom, Higgses and third generation of squarks are 
given in appendix~A, both at zero temperature and in the thermal plasma.
The $n_i$'s are the number of degrees of freedom and can also be found
in the appendix~A, $Q$ is the renormalization scale, and $C_i=5/6$ 
$(3/2)$ for gauge bosons (scalars and fermions). The mass parameters 
$m_{1,2}^2$ can be traded by $v_{1,2}^2$ by minimizing the $T=0$ 
one--loop potential \cite{beqz}. The $\widetilde{J}_i$'s give the 
finite temperature effects; the tilde denotes that daisy resummation 
has been performed in the $J_i$'s. Here $J_i=J_{\pm}$ is the free 
energy of an ideal gas of particles of mass $m_i(\varphi_1,\varphi_2)$,
\be
\label{ypsilon}
J_{\pm} (m^2/T^2) \equiv \int_0^{\infty} dx \, x^2 \,
\log \left( 1 \mp e^{- \sqrt{x^2 + m^2/T^2}} \right) \, ,
\ee
where $+(-)$ is for bosons (fermions). No Daisy resummation 
is needed for fermions and $\widetilde{J}_{-} =J_{-}$ . 
In the scalar sector, we choose to perform this resummation not only on
the zero Matsubara modes but on all of them \cite{par}, so that
\be
\widetilde{J}_{+}^{(scalar)} (m_i^2) = J_{+} (\ov{m}_i^2) \;\;,
\ee
where the $\ov{m}_i^2$ are the thermally corrected masses as given in 
appendix~A. We perform daisy resummation for gauge bosons by screening
the $n=0$ modes of longitudinal components $W_L$, $Z_L$, 
$\gamma_L$, $g_L$:
\be
\label{long}
\widetilde{J}_{+} (m_{i,long}^2) = J_{+} (m_i^2) - \frac{\pi}{6} 
\frac{m_{iL}^3-m_i^3}{T^3} \;\;,
\ee
where $m_{iL}$ are the Debye masses (to be found in appendix~A). 
In eq.~(\ref{long}) the $n=0$ (cubic) contribution to $J_+$ is 
subtracted out and replaced by a similar term with the thermally 
corrected mass. Transverse modes are not screened at leading order. 

This potential was studied in \cite{beqz}, where it was shown that the
presence of light stops can influence the strength of the electroweak
phase transition through their contribution to the cubic $n=0$ terms. 
The final region in parameter space where $v/T_c>1$ is 
determined by the interplay between two opposite effects; soft 
and thermal screening masses tend to decrease $v/T_c$ by screening the
pure cubic behaviour of $m_{\tilde t}^3$, while a large Yukawa coupling
tends to increase $v/T_c$. The numerical study of the potential shows 
that in some region of parameter space the cubic term from stops 
can dominate the electroweak phase transition.

However, the only region where this transition was strong enough for 
baryogenesis was limited to the large $m_A$, small $\tan \beta$, negligible 
stop mixing regime. For example, with a top quark pole mass $M_t=156\
GeV$,\footnote{We choose this too low value of $M_t$ for the sake of the
discussion as it corresponds roughly to the point were the effect of two-loop
corrections on $v/T_c$ is maximized. Results for $M_t=175\ GeV$ are presented
later.} stop soft masses $m_Q=70\ GeV$, $m_U=0$, zero stop mixing
$A_t=\mu=0$ and 
$\tan \beta=2.5$ the ratio $v/T_c$, which is $\sim 1$ for large $m_A$, 
drops below $0.5$ for $m_A=50\ GeV$. This is shown in curve $(b)$ 
of figure~2. Here $v=\sqrt{v_1^2+v_2^2}$, and $T_c$ is defined by the
coexistence of two degenerate minima in the T--dependent effective 
potential, one at $(\varphi_1,\varphi_2)=(0,0)$ and the other at 
$(v_1,v_2)$. The behaviour shown in figure~2 agrees well with the 
qualitative discussion of the previous section. Some comments on the 
choice of parameters are in order.
Unlike the case for large $m_A$, negligible stop mixing [$\sim 
(A_t\varphi_2-\mu\varphi_1)$] with two Higgs background fields 
necessarily implies negligible $A_t$ and $\mu$. Small $\mu$ may be 
problematic phenomenologically\footnote{We follow here the
conventional definition of $A_t$ and $\mu$, which differs from the more
general one used in \cite{laine}.}. However, as we have mentioned, 
the transition takes place for low values of $m_A$ along a field direction
for which $\varphi_2/\varphi_1$ is large. As a consequence, the
effect of $\mu$ in the off--diagonal stop mixing is suppressed and its
influence on the transition diminishes, which has been actually observed
in \cite{laine}. In practice, 
larger values of $\mu$ can be accommodated without weakening the
strength of the transition. We can then safely set $A_t=\mu=0$ for the
purpose of analyzing the transition strength and this simplifies 
considerably the analysis. In the following we refer to 
$\tilde{t}_R$ and $\tilde{t}_L$ as mass eigenstates.
\begin{figure}[hbt]
\centerline{
\psfig{figure=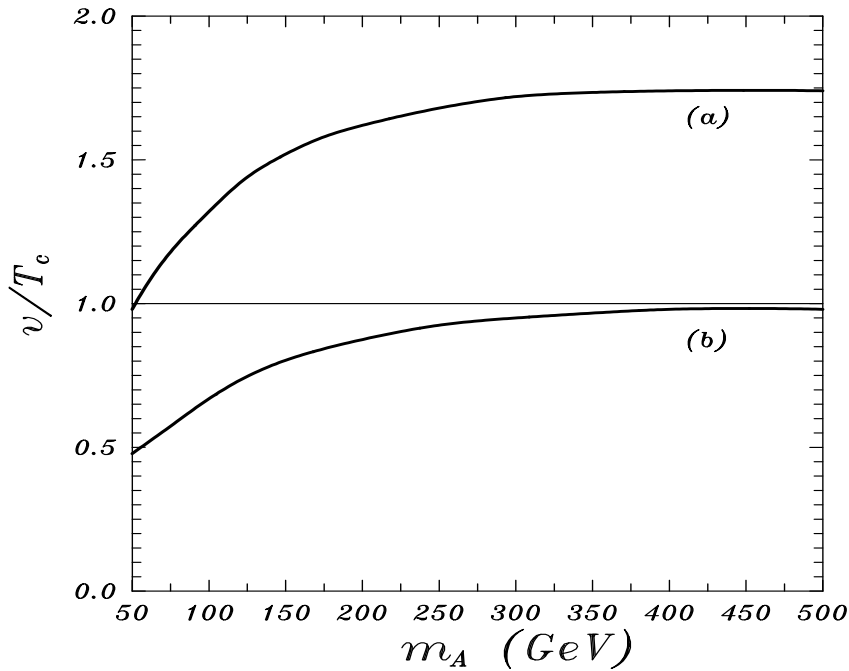,height=10cm,width=12cm,bbllx=2.cm,bblly=2.cm,bburx=15.cm,bbury=15cm}}
\caption{\footnotesize Order parameter $v/T$ at $T_c$ as a function of the
pseudoscalar mass $m_A$. Curve $(a)$ corresponds to the 2-loop resummed 
approximation, while $(b)$ gives the 1-loop resummed result.
Here $M_t=156\ GeV$, $m_Q=70\ GeV$, $m_U=0$ and $\tan\beta=2.5$.} 
\end{figure} 

Values of $m_Q$ lower than $70\ GeV$ would increase $v/T_c$, but 
they would be in conflict with $\Delta\rho$ constraints \cite{eqz}. 
Another possibility to increase $v/T_c$ is to choose 
$m_U^2<0$ \cite{cqw}. In that case, $\tilde{t}_R$ may get a vev and so has
to be included in
the discussion of the effective potential. Although, if realized, this
region of parameters could have important implications (see 
\cite{cqw,bojolasc}) we do not explore it in this paper and we stop at
$m_U=0$. The perturbative analysis of the $T\neq 0$ potential along
the squark direction (necessary to ensure that $m_U^2<0$ is allowed)
may be problematic. We can make a rough estimate of the expansion 
parameter in the colour breaking minimum that may develop in the
squark direction; in analogy with the SM, this expansion 
parameter would be the ratio of the squark quartic coupling over the 
$SU(3)$ gauge coupling squared. This ratio is in principle of 
order one. In addition, when the thermally corrected mass of $\tilde{t}_R$
is close to zero, radiative corrections can grow very large and affect
the convergence of the perturbation series along the Higgs 
direction. These problems have been observed in ref.~\cite{bojolasc}
where a two-loop analysis of the potential (both along the Higgs and stop
directions) was presented for the case $m_U^2<0$.  Clearly a non-perturbative
analysis of this region ($m_U^2<0$) would be desirable.

However, sufficiently large values of $v/T_c$ are obtained when 
two--loop corrections are taken into account \cite{esp}, as we 
discuss in the next section, so that one is not confined to the 
choice $m_U^2<0$. In particular, this means that the condition 
$m_{\tilde{t}_1}<m_t$ is {\em not} required for a sufficiently strong 
phase transition.

\vspace{.5cm}
\section{Two-loop resummed potential} 

For simplicity, we set $g'=0=h_b$ in two--loop corrections. We only 
consider the case of negligible stop mixing, as this is the best 
case for a strong phase transition \cite{beqz,laine}. 
 
\begin{figure}[hbt]
\centerline{
\psfig{figure=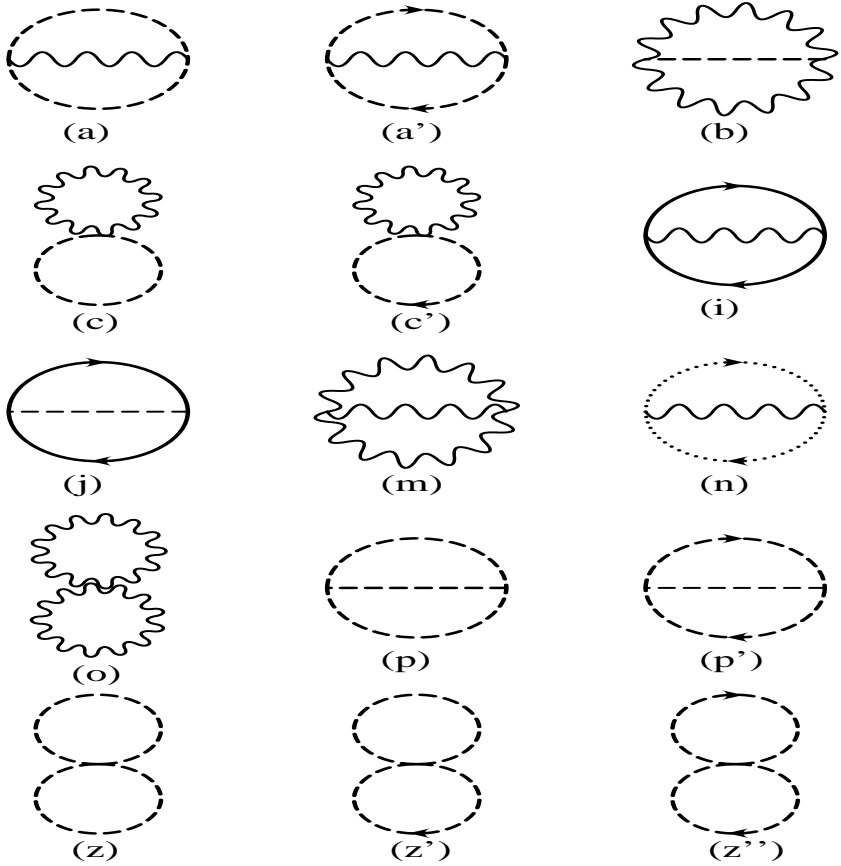,height=12cm,width=10cm,bbllx=5.cm,bblly=2.cm,bburx=17.cm,bbury=28cm}}
\caption{\footnotesize Two-loop vacuum graphs contributing to the 
effective potential (in Landau gauge). Dashed lines represent Higgs 
bosons, wiggled lines gauge bosons, dotted lines ghosts. Dashed lines 
with an arrow represent squarks, while quarks are solid lines with an 
arrow.} 
\end{figure} 

The two--loop diagrams to consider are displayed in figure~3. 
Counterterm graphs are not shown. We represent Higgs bosons ($A^0$, 
$H^0$, $h^0$, $G^0$, $G^{\pm}$, $H^{\pm}$) by simple dashed lines, 
gauge bosons (g, W, Z, $\gamma$) by wiggly lines, squarks 
($\tilde{t}_{L,R}$, $\tilde{b}_{L,R}$) by dashed lines with arrow, 
quarks (t,b) by continuous lines with arrow, and ghosts ($c_W$, $c_Z$) 
by dotted lines with arrow. Within the stated approximations, the full 
two--loop corrections to the effective potential are collected in 
appendix~B (we follow the notation of
refs.~\cite{arnes,twolsm}). In the numerical analysis we use a high 
temperature expansion of the different integrals, keeping terms up to 
order $g_i^4$, $h_t^4$. In some corners of the parameter space we 
explore, for example the regions with large values of $m_A$ and/or $m_Q$,
this high temperature expansion starts to fail. We control
our numerical results in those cases interpolating between the region
where the expansion
is safe and the large $m_A, m_Q$ region where the $T\neq 0$ 
contributions of the heavy Higgs  and/or $(\tilde{t},\tilde{b})_L$ 
doublet are dropped.

The dominant two-loop terms, i.e., those that have a greater influence 
on the transition strength, are written down in appendix C. These are 
terms involving logarithms of masses. In some cases, these 
logarithmic terms introduce a linear dependence on the fields 
$\varphi_1$ and/or $\varphi_2$ for small $\varphi_{1,2}$ which is 
cancelled by other terms in the potential. We keep these terms  
to ensure that our approximation is well behaved in that respect. 
Finally, we also add the non--logarithmic terms which depend on 
the couplings $g_s$ and $h_t$ and thus can be potentially large 
(although they mainly affect the transition temperature and do not 
change $v/T_c$ too much). 

The relative importance of the different diagrams follows the pattern 
discussed in ref.~\cite{esp} for the large $m_A$ case. For low values 
of the stop soft masses, the main effect is due to diagram $(a')$ with 
gluon exchange. This QCD contribution significantly increases 
 the value of $v/T$ at the transition. This important enhancement is
operative also in the low $m_A$ regime as shown in fig.~2.
Curve $(a)$ gives $v/T_c$ computed from the 
two--loop resummed potential as a function of the pseudoscalar mass 
$m_A$. The corresponding one-loop result is given by curve $(b)$. The 
parameters have been chosen to maximize the effect.

\vspace{.5cm}
\section{Results} 

Before presenting the results of our numerical computation of the 
transition strength some comments on our definition of $v/T_c$ are in 
order. We have taken $T_c$ as the temperature of degeneracy between 
the symmetric and broken--phase minima. A possible alternative is to 
choose the temperature $T_0$ at which the symmetric minimum gets 
destabilized (that is, $V''(0)=0$ along some field direction); 
in that case, the necessary condition 
for baryogenesis is $v/T_0\simgt 1.2-1.3$. One can
refine the precise number entering this condition by estimating the 
sphaleron energy \cite{beqz,moq}. The true critical 
temperature should be in the narrow interval $[T_0,T_c]$ if the 
transition is first-order. When two--loop corrections
are included, in particular those of the form $\delta V\sim 
T^2\varphi^2\log\varphi$, $V''(\varphi=0)$ diverges.
This behaviour is associated with the (non-screened) transverse modes 
of gauge bosons. In principle, this prevents the
computation of $T_0$. For practical purposes, one can 
overcome this difficulty by turning on a small magnetic mass (which, 
after all, is generated non--perturbatively\footnote{In the
abelian--Higgs model no 
such magnetic mass is generated, even non--perturbatively. However, 
the diagrams responsible for the divergence are also absent as was to 
be expected.}) and checking that $T_0$ is not sensitive to its 
particular value. Although we have chosen to work with 
$T_c$, which can be defined in a cleaner way, we checked that the 
results obtained for $v/T_0$ are consistent with those that we 
present now.

\begin{figure}[hbt]
\centerline{
\psfig{figure=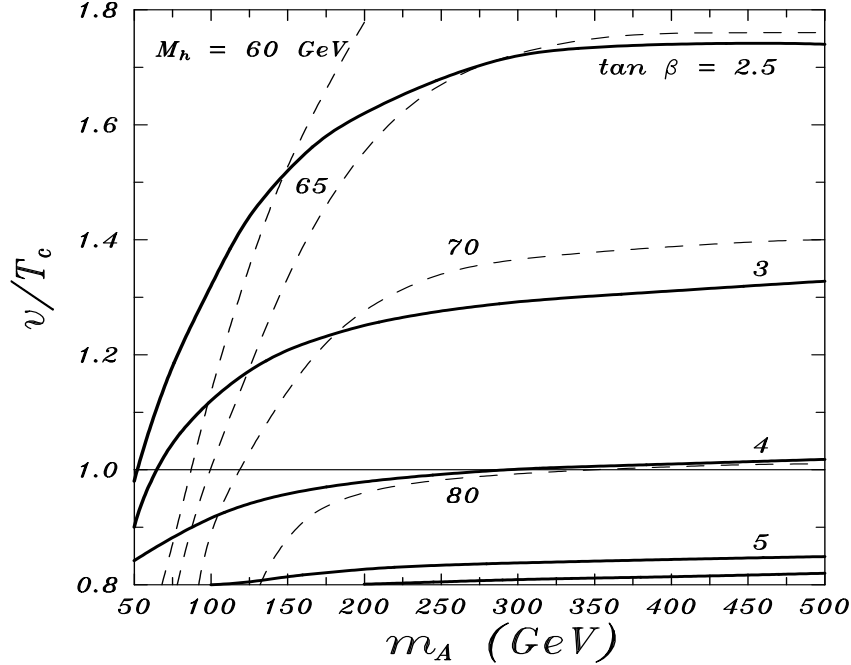,height=10cm,width=12cm,bbllx=2.cm,bblly=2.cm,bburx=15.cm,bbury=15cm}}
\caption{\footnotesize Lines of $v/T_c$ (solid) for different values of 
$\tan\beta$ as a function of $m_A$. The mass of the lightest
$CP$--even Higgs is given by the dashed contour lines. $M_t=156\ GeV$,
$m_U=0$, $m_Q=70\ GeV$.} 
\end{figure} 
\begin{figure}[hbt]
\centerline{
\psfig{figure=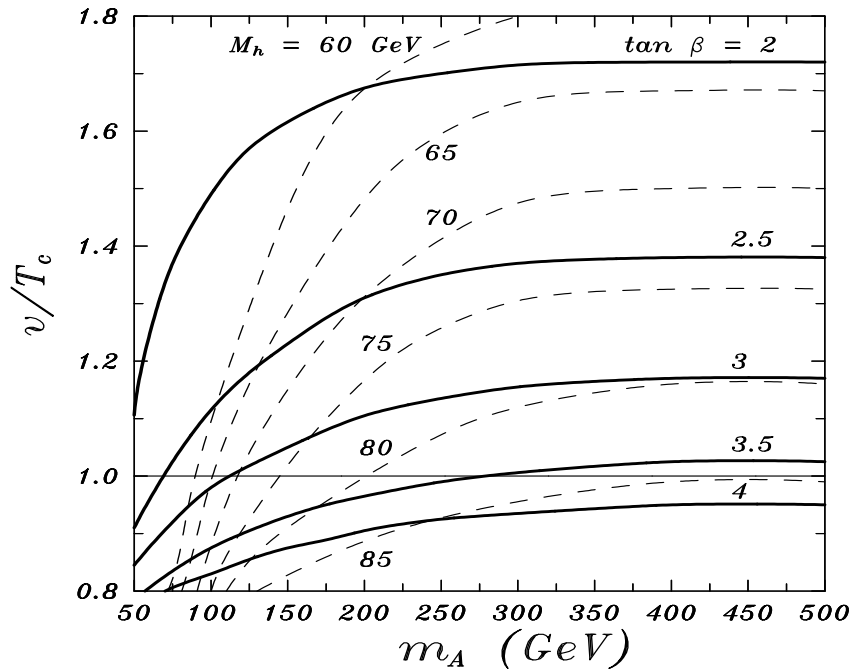,height=10cm,width=12cm,bbllx=2.cm,bblly=2.cm,bburx=15.cm,bbury=15cm}}
\caption{\footnotesize 
Same as fig.~4 but for $M_t=175\ GeV$, $m_U=0$, $m_Q=250\ GeV$.} 
\end{figure} 

In figure 4 we plot $v/T_c$ (solid lines) as a function of the 
pseudoscalar mass $m_A$ for different values of $\tan\beta$. The other
relevant parameters are $M_t=156\ GeV$ (a low value that roughly 
corresponds to the maximal effect of two--loop corrections), $m_U=0$, 
and $m_Q=70\ GeV$ (to satisfy the $\Delta\rho$ constraint). In 
figure 5 we repeat the plot for $M_t=175\ GeV$ ($m_U=0$, $m_Q=250\ 
GeV$). These plots show that the transition is always stronger for 
lower values of $\tan\beta$ and larger values of $m_A$. At large
$m_A$, $v/T_c$ gets stabilized and the results of ref.~\cite{esp} are 
recovered\footnote{The apparent discrepancy between figure~5 at large $m_A$ and
the results of \cite{esp} for the same top mass is due to a more restrictive
choice of $m_Q$ in \cite{esp}. There, a one-loop expression for $\Delta\rho$
was used to estimate the lower limit on $m_Q$. This limit is weaker when $QCD$
corrections are included (see \cite{djouadi} and references therein).}.
At low $m_A$, $v/T_c$ drops and the curves for
different values of $\tan\beta$ get focused in a narrow range. This behaviour, 
previously observed in refs.~\cite{laine,falo}, has been discussed 
in section~3. For a given value of $\tan\beta$,
there is a critical value of $m_A$ below which the transition is too 
weak for baryogenesis. This critical value of $m_A$ increases with
increasing $\tan\beta$ and goes to infinity for
$\tan\beta\sim 3.5-4$. For larger $\tan\beta$ the
transition is too weak for any value of $m_A$.

Small $\tan\beta$ values ($<2$) give a very strong transition because 
the Higgs mass $m_{h}$ is small. Dashed lines are contour lines
for $m_{h}$. In the way we are plotting our results, these curves 
are determined by the solid ones, because $m_h$ depends on $m_A$
and $\tan\beta$. We use the LEP limits on $m_{h}$ to determine the
experimentally allowed region. For large $m_A$, these limits arise 
from the non-observation of the production process $Z\rightarrow 
h^0Z^*$, and we take the Standard Model lower limit $m_{h^0}>70.7\ GeV$, 
because in that regime $h^0$ has SM couplings. For light $A^0$, the 
process $Z\rightarrow h^0A^0$ becomes important and in fact sets the 
strongest limit in the low $m_A$--large $\tan\beta$ region. However, 
that region already corresponds to $v/T_c<1$. For our purposes,
we can take the line of $m_{h^0}=70\ GeV$ as the limit of the allowed 
region (it gets a few $GeV$ weaker when $m_A\sim 100\
GeV$). After imposing the LEP bound we find from fig.~5 an absolute 
lower bound on $m_A$ of about $120\ GeV$ (below which the transition
is too weak for baryogenesis)\footnote{For low values of $m_A$, the charged Higgs
boson is also light and
gives a large contribution to $b\rightarrow s\gamma$ which may conflict
with the measured value of this branching ratio. 
However, cancellations with other supersymmetric
contributions (from light stops and charginos) may occur and
we cannot exclude these low values of $m_A$ without further assumptions
on the SUSY parameters.} which corresponds to 
$\tan\beta\sim 3$. This lower bound grows for larger $\tan\beta$. The
upper limit on $\tan\beta$ is $\tan\beta\simlt 3.6$ and the lower limit is
$\tan\beta\simgt 2.25$. The 
Higgs mass is below $85\ GeV$, within the reach of LEP II. We find no trace of
a good region for baryogenesis at low $m_A$ and large $\tan\beta$.

Our results are in good agreement with those obtained in 
ref.~\cite{laine} using a 3d reduced effective theory and lattice 
Monte Carlo results. The upper bound on $m_h$ of about $70\ GeV$
quoted in that paper is smaller than our bound of $85\ GeV$ because of
the choice $m_U=100\ GeV$ in \cite{laine}. Smaller values of $m_U$
would allow for larger Higgs masses, as is shown in figure~5 of that 
paper, but in that region the approximations used to construct the 
reduced 3d theory start to break down. The same comments apply to the 
lower bound on $m_A$ of $m_A\simgt 200\ GeV$, which can be lowered for 
smaller values of $m_U$.

An independent 3d analysis of the transition has been carried out in 
refs.~\cite{falo}. It is more difficult to compare our results
to theirs because they imposed SUGRA--type constraints on the parameters
of the model. Of course, this restriction leads to stronger
bounds on $m_h$, $m_A$ and $\tan\beta$ but the qualitative dependence of
the transition strength on different parameters is similar to our
results.

\vspace{.5cm}
\section{Conclusion}

In this paper, we have studied the electroweak phase transition in the 
MSSM, searching for regions where it is strong enough to permit 
electroweak baryogenesis. If stops are only moderately heavy they 
control the strength of the transition and the use of perturbation 
theory for its study is well justified. We 
first addressed the question of whether the electroweak phase 
transition can be a crossover in the MSSM. We clarified the 
differences with respect to the Standard Model case and estimated the 
critical Higgs mass for crossover, which depends on how stops affect 
the transition. This critical Higgs mass (beyond which the analytic 
crossover can take place) can be significantly larger than in the 
Standard Model.

We scanned the parameter space $(m_A,\tan\beta)$, 
qualitatively explaining the influence on the transition strength
expected from different effects. We identified a region with light 
(unmixed) stops, light $m_h$, and small $\tan\beta$ where the transition 
can be strong. That region is widest for large $m_A$, and shrinks when
$m_A$ is lowered. We then 
focused on the interesting low  $m_A$ region and performed a
numerical study evaluating the $T$-dependent effective 
potential of the model (simplified to a two Higgs doublet model plus 
third generation squarks) up to two--loop resummed contributions. We 
found that the region where the transition is strong extends down to 
$m_A\sim120\ GeV$ with roughly $2\simlt \tan\beta\simlt 4$ and $m_h\simlt 85\ GeV$ 
(light stops are required but $m_{\tilde{t}}<m_t$ is {\em not} a 
necessary condition). This region will be explored by LEP II in the 
near future, which will stringently test the viability of
electroweak baryogenesis in the MSSM.

\vspace{0.5cm}

{\bf Acknowledgements}

We thank J. Cline and K. Kainulainen for kindly providing us with some
of the data for the analysis of ref.~\cite{clka}. We also thank L.~Everett
for useful suggestions. B. de C. thanks the Physics and Astronomy Dept.
of the University of Pennsylvania and J.R.E. thanks the Centre for
Theoretical Physics of the University of Sussex for hospitality during
some stages of this work, which was  supported by PPARC (B. de C.) and the
U.S. Department of Energy Grant No. DOE-EY-76-02-3071 (J.R.E.). 

\newpage

\renewcommand{\theequation}{\thesection.\arabic{equation}}
\catcode`\@=11
\@addtoreset{equation}{section}

\appendix

\vspace{0.5cm}
\section{Field-dependent Masses and Mixing Angles}

We give here the field--dependent masses (in the background
$\varphi_{1,2}=\langle H_{1,2}^0\rangle\sqrt{2}$) of different species
of particles relevant for the effective potential. Where appropriate,
field--dependent mixing angles are defined. We also give the leading 
parts of thermal masses, needed for resummation of the potential. 
These masses are computed \cite{esco} assuming that the plasma is 
populated by SM particles, an extra Higgs doublet and squarks of the 
third generation.

\begin{itemize}
\item Gauge bosons:

The number of degrees of freedom is $n_W=6$ (with $n_{W_L}=2, n_{W_T}=4$)
and $n_{Z,\gamma}=3$ $(n_{Z_L,\gamma_L}=1, n_{Z_T,\gamma_T}=2)$. 
The masses are
\begin{equation}
M_W^2 = \frac{1}{4} g^2 (\varphi_1^2 + \varphi_2^2) \; ,\;\;\;\;
M_{W_3-B}^2 = \frac{1}{4}
\left(
\begin{array}{cc}
g^2  & -gg' \\
-gg' &  g'^2
\end{array}
\right)
(\varphi_1^2 + \varphi_2^2),
\end{equation}
which gives
\be
M_Z^2=\frac{1}{4}G^2(\varphi_1^2+\varphi_2^2)\;,\;\;\;\;
M_\gamma^2=0,
\ee
where $G^2=g^2+g'^2$.

At finite temperature the leading contribution to the thermal mass is
zero (at leading order $gT$ and to all orders in perturbation theory) 
for transverse modes while for longitudinal modes the above 
expressions change to
\bear
M_{W_L}^2&\equiv &M_W^2+ \Pi_{W_L}=M_W^2+\frac{5}{2}g^2T^2,\nonumber\\
M_{(W_3-B)_L}^2&\equiv & M_{W_3-B}^2+\left(
\begin{array}{cc}
\Pi_{W_L} & 0 \\
0 & \Pi_{B_L}
\end{array}
\right),
\eear
with $\Pi_{B_L}=\frac{47}{18}g'^2T^2$. The eigenvalues of 
$M_{(W_3-B)_L}^2$ are $M_{Z_L}^2$ and $M^2_{\gamma_L}$. In the 
approximation $g'=0$ (used for the two-loop corrections in this paper)
it follows that $M_{\gamma_L}=0$ and $M_{Z_L}=M_{W_L}\equiv M_L$.

For gluons ($n_g=3\times 8$) the masses are
\be
M_{g_T}^2=0 \;,\; \;\;\; M_{g_L}^2=0+\Pi_{gL}=\frac{8}{3}g_s^2T^2.
\ee

\item Quarks: 

For the third generation we have ($n_t=n_b=-12$)
\begin{equation}
m_t = \frac{1}{\sqrt{2}} h_t \varphi_2 ,\;\;\;\;
m_b = \frac{1}{\sqrt{2}} h_b \varphi_1 . 
\end{equation}
where $h_t$ ($h_b$) is the top (bottom) Yukawa coupling.

\item Squarks: 

We consider only the third generation, that is, stops and sbottoms 
(with $n_{\tilde{t}_L}=n_{\tilde{t}_R}=n_{\tilde{b}_L}=n_{\tilde{b}_R}
=6$). The mass eigenstates are given by the diagonalization of 
their $2\times 2$ mass matrix defined in terms of interaction 
eigenstates. For stops this is given by:
\begin{eqnarray}
\left(\begin{array}{cc}
m^2_{Q} + m_t^2 + \frac{1}{24} (3 g^2-g'^2) (\varphi_1^2 - 
\varphi_2^2) & 
\frac{1}{\sqrt{2}} h_t (A_t \varphi_2+ \mu \varphi_1) \\
\frac{1}{\sqrt{2}} h_t (A_t \varphi_2+ \mu \varphi_1) & 
m^2_{U} + m_t^2 + \frac{1}{6} g'^2 (\varphi_1^2 - \varphi_2^2)  
\end{array}\right), \nonumber
\end{eqnarray}
while for sbottoms it is:
\begin{eqnarray}
\left(\begin{array}{cc}
m^2_{Q} + m_b^2 - \frac{1}{24} (3 g^2+g'^2) (\varphi_1^2 - 
\varphi_2^2) & 
\frac{1}{\sqrt{2}} h_b (A_b \varphi_1 + \mu \varphi_2) \\
\frac{1}{\sqrt{2}} h_b (A_b \varphi_1 + \mu \varphi_2) & 
m^2_{D} + m_b^2 - \frac{1}{12} g'^2 (\varphi_1^2 - \varphi_2^2) 
\end{array}\right) .\nonumber
\end{eqnarray}  
At finite $T$, the masses corrected by thermal effects are given (at
leading order) by
\be
\overline{M}^2_{\tilde t}=M^2_{\tilde t}+
\left(
\begin{array}{cc}
\Pi_Q & 0 \\
0 & \Pi_U
\end{array}
\right)
\;\; ; \;\;
\overline{M}^2_{\tilde b}=M^2_{\tilde b}+
\left(
\begin{array}{cc}
\Pi_Q & 0 \\
0 & \Pi_D
\end{array}
\right),
\ee
with
\bear
\Pi_Q&=&\frac{4}{9}g_s^2T^2+\frac{1}{4}g^2T^2+\frac{1}{108}g'^2T^2
+\frac{1}{6}h_t^2T^2,\nonumber\\
\Pi_U&=&\frac{4}{9}g_s^2T^2+\frac{4}{27}g'^2T^2+\frac{1}{3}h_t^2T^2,
\nonumber\\
\Pi_D&=&\frac{4}{9}g_s^2T^2+\frac{1}{27}g'^2T^2+\frac{1}{3}h_t^2T^2.
\nonumber
\eear
Mixing angles can be defined in the usual way. As we are interested 
in the case of negligible squark mixing we do not give expressions for
these angles explicitly. In this case the mass and interaction eigenstates 
coincide.

\item Higgs sector: 

The two Higgs doublets are 
\begin{equation}
\left(\begin{array}{c}
\frac{1}{\sqrt{2}} (\hu+\varphi_1+i h_1^{0i}) \\ H_1^{-} 
\end{array}\right) , \;\; \; \;\;
\left(\begin{array}{c}
H_2^+ \\ \frac{1}{\sqrt{2}} (h_2^{0r}+\varphi_2+i h_2^{0i}) 
\end{array}\right) \;\; . 
\end{equation}
The background $\varphi_{1,2}$ is $CP$ conserving so that the scalar 
$(h_1^{0r},h_2^{0r})$, pseudoscalar $(h_1^{0i},h_2^{0i})$ and
charged $(H_1^{-*},H_2^+)$ sectors can be treated separately. For each 
of them we have a $2\times2$ hermitian mass matrix 
\begin{eqnarray}
\label{massmatrix}
{{\cal M}_{(k)}}^2=
\left(\begin{array}{cc}
{M^{(k)}}_{11}^{2} & {M^{(k)}}_{12}^{2} \\
{M^{(k)}}_{21}^{2} & {M^{(k)}}_{22}^{2}
\end{array}\right) ,
\end{eqnarray}
(where $k=r,i,c$) and a mixing angle $\theta_k$ which relates mass 
eigenstates ($H,h$) to interaction eigenstates ($H_1,H_2$) according to
\begin{eqnarray}
\left(
\begin{array}{c}
H_1^{(k)} \\
H_2^{(k)}
\end{array}
\right)
=
\left(
\begin{array}{cc}
\cos\theta_k & -\sin\theta_k \\
\sin\theta_k & \cos\theta_k
\end{array}
\right)
\left(
\begin{array}{c}
H^{(k)} \\
h^{(k)}
\end{array}
\right).
\end{eqnarray}
The mixing angles are then given by:
\begin{eqnarray}
\sin 2\theta_k & = & \frac{2 {{M}^{(k)}}_{12}^{2}}
{\sqrt{({{M}^{(k)}}_{11}^{2}-{{M}^{(k)}}_{22}^{2})^2
+4 ({{M}^{(k)}}_{12}^{2})^2}}, \nonumber \\
\label{angles} \\
\cos 2\theta_k & = & \frac{{{M}^{(k)}}_{11}^{2}-
{{M}^{(k)}}_{22}^{2}}
{\sqrt{({{M}^{(k)}}_{11}^{2}-{{ M}^{(k)}}_{22}^{2})^2
+4 ({{M}^{(k)}}_{12}^{2})^2}}\;\; . \nonumber
\end{eqnarray}
For the $CP$-even sector, $H_{1,2}^{(r)}=h_{1,2}^{0r}$, 
$h^{(r)}\equiv h^0$, $H^{(r)}\equiv H^0$, $n_{h^0}=n_{H^0}=1$ and 
(\ref{massmatrix}) reads:
\begin{equation}
\label{massr}
{{\cal M}_{(r)}}^{2} = 
\left(
\begin{array}{cc}
 m_1^2 + \frac{G^2}{8} (3 \varphi_1^2 - \varphi_2^2 ) &
 m_{12}^2 - \frac{G^2}{4} \varphi_1 \varphi_2 \\  
 m_{12}^2 - \frac{G^2}{4} \varphi_1 \varphi_2  &
 m_2^2 + \frac{G^2}{8} (3 \varphi_2^2 - \varphi_1^2 )
\end{array}
\right),
\end{equation}
and, at finite T,
\be
\label{masst}
{\overline{\cal M}_{(r)}}^{2} ={{\cal M}_{(r)}}^{2}+
\left(
\begin{array}{cc}
\Pi_1 & 0 \\
0 & \Pi_2
\end{array}
\right) ,
\ee
with
\be
\Pi_1=\frac{1}{4}g^2T^2+\frac{1}{12}g'^2T^2\;,\;\;\;\;
 \Pi_2=\frac{1}{4}g^2T^2+\frac{1}{12}g'^2T^2+\frac{3}{4}h_t^2T^2.
\ee
There is a mixing angle $\theta_r$ for the diagonalization of 
(\ref{massr}) and a T--dependent angle $\overline{\theta}_r$ for the 
diagonalization of (\ref{masst}).

For the $CP$-odd sector, $H_{1,2}^{(i)}=h_{1,2}^{0i}$, $h^{(i)}\equiv 
G^0$, $H^{(i)}\equiv A^0$, $n_{G^0}=n_{A^0}=1$ and (\ref{massmatrix}) 
reads:
\begin{equation}
{{\cal M}_{(i)}}^{2} = 
\left(
\begin{array}{cc}
 m_1^2 + \frac{G^2}{8} ( \varphi_1^2 - \varphi_2^2 ) & -m_{12}^2  \\  
 -m_{12}^2  & m_2^2 + \frac{G^2}{8} (\varphi_2^2 - \varphi_1^2 )
\end{array}
\right),
\end{equation}
and, at finite T, an equation similar to (\ref{masst}) holds, 
with $r\rightarrow i$.

For the charged sector, $H_1^{(c)}=H_{1}^{-*}$, $H_2^{(c)}=H_2^+$, 
$h^{(c)}\equiv G^+$, $H^{(c)}\equiv H^+$, $n_{G^\pm}=n_{H^\pm}=2$ 
and (\ref{massmatrix}) reads:
\begin{equation}
{{\cal M}^{(c)}}^{2} = 
\left(
\begin{array}{cc}
 m_1^2 + \frac{G^2}{8}\varphi_1^2 +\frac{(g^2-g'^2)}{8} \varphi_2^2  
 & -m_{12}^2 +\frac{g^2}{4}\varphi_1\varphi_2 \\  
 -m_{12}^2 +\frac{g^2}{4}\varphi_1\varphi_2 
  & m_2^2 + \frac{G^2}{8}\varphi_2^2+\frac{(g^2-g'^2)}{8}  \varphi_1^2 
\end{array}
\right),
\end{equation}
and, at finite T, an equation similar to (\ref{masst}) holds, 
with $r\rightarrow c$.

For different cross-checking purposes it is useful to consider the 
behaviour of the mixing angles $\theta_k$ in two different limits:

{\em (a)} In the large $m_A$ limit
\be
\theta_r\rightarrow \beta + \frac{\pi}{2}\;, \;\;\;\;
\theta_i,\theta_c\rightarrow -\beta - \frac{\pi}{2}.
\ee

{\em (b)} In the physical limit, $\varphi_{1,2}\rightarrow v_{1,2}$,
\be
\theta_r\rightarrow \alpha \;,\;\;\;\;
\theta_i,\theta_c\rightarrow -\beta+\frac{\pi}{2},
\ee
where $\alpha$ is the physical mixing angle in the $CP$-even sector.

For later use it is convenient to define the following abbreviations 
(with the indicated limiting values $[{\em (a),(b)}]$):
\be
\begin{array}{lcl}
\varphi_0\equiv\varphi_1\cos\theta_r+\varphi_2\sin\theta_r &
\rightarrow & [0,v\cos(\beta-\alpha)]\nonumber\\
\varphi_v\equiv\varphi_2\cos\theta_r-\varphi_1\sin\theta_r &
\rightarrow & [\varphi,v\sin(\beta-\alpha)]\nonumber\\
\varphi_{sin}\equiv\varphi_1\cos\theta_r-\varphi_2\sin\theta_r &
\rightarrow & [\varphi\sin 2\beta,v\cos(\beta+\alpha)]\nonumber\\
\varphi_{cos}\equiv\varphi_2\cos\theta_r+\varphi_1\sin\theta_r &
\rightarrow & [\varphi\cos 2\beta,v\sin(\beta+\alpha)]\nonumber\\
\varphi_{0,i}\equiv\varphi_2\sin\theta_i-\varphi_1\cos\theta_i &
\rightarrow & [0,0]\nonumber\\
\varphi_{v,i}\equiv\varphi_2\cos\theta_i+\varphi_1\sin\theta_i &
\rightarrow & [\varphi,v]\nonumber\\
\varphi_{cos,c}\equiv\varphi_2\cos\theta_c-\varphi_1\sin\theta_c &
\rightarrow & [-\varphi\cos 2\beta,-v\cos 2\beta]\nonumber\\
\varphi_{sin,c}\equiv\varphi_2\sin\theta_c+\varphi_1\cos\theta_c &
\rightarrow & [\varphi\sin 2\beta,v\sin 2\beta].\nonumber
\end{array}
\ee
\end{itemize}
\vspace{0.5cm}

\section{Two-loop Resummed Potential}

We present the resummed two-loop corrections to the finite 
temperature effective potential in the MSSM framework described in 
the text. We make the approximation $g'=h_b=0$, so that in the 
formulae below we make no distinction between $m_Z$ and $m_W$ which 
are called $M$ ($M_L$ is the longitudinal component). We further 
assume that left-right mixing in the squark sector is small and can be 
neglected. We follow the notation of \cite{arnes,twolsm} and the labeling of
different contributions corresponds to our figure~3. As usual,
$N_c=3$ counts the number of colours and we use the abbreviations 
$c_r=\cos\theta_r$, $s_{2c}=\sin 2\theta_c$, etc. For some recurrent 
combinations of fields and angles (like e.g. $\varphi_1 c_r+\varphi_2 
s_r$) we use the short-hand expressions defined at the end of appendix~A. 
With our resummation method in the scalar sector (we included the thermal mass 
correction in all Matsubara modes) masses and mixing angles are 
$T$--dependent: $\overline{m}_i$, $\overline{\theta}_k$. For
simplicity we drop the bars in the formulae below and simply write 
$m_i$, $\theta_k$ everywhere.
\bear
V^{(a)}=&-&\frac{g^2}{8}\left\{
2\cos^2(\theta_r+\theta_c)[
{\cal D}_{SSV}(m_{G^\pm},m_{h^0},M)
+{\cal D}_{SSV}(m_{H^\pm},m_{H^0},M)]\right.
\nonumber\\
&+&
2\sin^2(\theta_r+\theta_c)[ 
{\cal D}_{SSV}(m_{G^\pm},m_{H^0},M)+{\cal D}_{SSV}(m_{H^\pm},m_{h^0},M)]
\nonumber\\
&+&
2\cos^2(\theta_c-\theta_i)[ 
{\cal D}_{SSV}(m_{G^\pm},m_{G^0},M)+{\cal D}_{SSV}(m_{H^\pm},m_{A^0},M)]
\nonumber\\
&+&
2\sin^2(\theta_c-\theta_i)[
{\cal D}_{SSV}(m_{G^\pm},m_{A^0},M)+{\cal D}_{SSV}(m_{H^\pm},m_{G^0},M)]       
\nonumber\\
&+&
\cos^2(\theta_i+\theta_r)[
{\cal D}_{SSV}(m_{G^0},m_{h^0},M)+{\cal D}_{SSV}(m_{A^0},m_{H^0},M)]
\nonumber\\
&+&
\sin^2(\theta_i+\theta_r)[
{\cal D}_{SSV}(m_{G^0},m_{H^0},M)+{\cal D}_{SSV}(m_{h^0},m_{A^0},M)]
\nonumber\\
&+&\left.
{\cal D}_{SSV}(m_{G^\pm},m_{G^\pm},M)
+{\cal D}_{SSV}(m_{H^\pm},m_{H^\pm},M)]\right\},\nonumber\\
V^{(a')}=&-&\frac{g^2}{8}N_c\left[{\cal D}_{SSV}(\mtl,\mtl,M)+
{\cal D}_{SSV}(\mbl,\mbl,M)+
4 {\cal D}_{SSV}(\mtl,\mbl,M)
\right]\nonumber\\
&-&\frac{g_s^2}{4}(N_c^2-1)\left[{\cal D}_{SSV}(\mtl,\mtl,0)+
{\cal D}_{SSV}(\mbl,\mbl,0)\right.\nonumber\\
&+&\left.{\cal D}_{SSV}(\mtr,\mtr,0)+
{\cal D}_{SSV}(\mbr,\mbr,0)
\right],\nonumber\\
V^{(b)}=&-&\frac{3}{64}g^4\left[
\varphi_v^2{\cal D}_{SVV}(m_{h^0},M,M)+
\varphi_0^2{\cal D}_{SVV}(m_{H^0},M,M)]\right] ,
\nonumber\\
V^{(c)}=&-&\frac{3}{16}g^2\left[
2{\cal D}_{SV}(m_{G^\pm},M)+{\cal D}_{SV}(m_{G^0},M)+
{\cal D}_{SV}(m_{h^0},M)\right.\nonumber\\
&+&\left.
2{\cal D}_{SV}(m_{H^\pm},M)+{\cal D}_{SV}(m_{A^0},M)+
{\cal D}_{SV}(m_{H^0},M)\right],\nonumber\\
V^{(c')}=&-&\frac{1}{4}g_s^2(N_c^2-1)\left[{\cal D}_{SV}(\mtl,0)
+{\cal D}_{SV}(\mbl,0)+{\cal D}_{SV}(\mtr,0)+{\cal D}_{SV}(\mbr,0)
\right]\nonumber\\
&-&\frac{3}{8} g^2 N_c\left[{\cal D}_{SV}(\mtl,M)+
{\cal D}_{SV}(\mbl,M)
\right],\nonumber\\
V^{(i)}=&-&4g_s^2{\cal D}_{ffV}(m_t,m_t,0)-\frac{3}{8}g^2[
{\cal D}_{ffV}(m_t,m_t,M)-{\cal D}_{ffV}(0,0,M)]\nonumber\\
&-&\frac{3}{2}g^2[ {\cal D}_{ffV}(m_t,0,M)-{\cal D}_{ffV}(0,0,M)]
-3g^2n_t{\cal D}_{ffV}(0,0,M),\nonumber\\
V^{(j)}=&-&\frac{N_c}{4}h_t^2\left[
c_r^2{\cal D}_{ffS}(m_t,m_t,m_{h^0})+
s_r^2{\cal D}_{ffS}(m_t,m_t,m_{H^0})+
c_i^2{\cal D}_{ffS}(m_t,m_t,m_{G^0})
\right.\nonumber\\
&+&\left.s_i^2{\cal D}_{ffS}(m_t,m_t,m_{A^0})+
2c_c^2{\cal D}_{ffS}(m_t,m_b,m_{G^\pm})+
2s_c^2{\cal D}_{ffS}(m_t,m_b,m_{H^\pm})
\right],\nonumber\\
V^{(m)}=&-&\frac{1}{2}g^2\left[{\cal D}_{VVV}(M,M,M)+3
{\cal D}_{LLT}(M_L,M_L,M)-3{\cal D}_{LLT}(M,M,M) \right], \nonumber \\
V^{(n)}=&-&3g^2{\cal D}_{\eta \eta V}(M), \nonumber \\
V^{(o)}=&-&\frac{3}{4}g^2{\cal D}_{VV}(M,M), \nonumber \\
V^{(p)}=&-&\frac{1}{64}g^4\left\{
2\left[(\varphi_v-s_{2c}\varphi_{sin})^2
{\overline {\rm H}}(m_{h^0},m_{G^\pm},m_{G^\pm})
+(\varphi_v+s_{2c}\varphi_{sin})^2
{\overline {\rm H}}(m_{h^0},m_{H^\pm},m_{H^\pm})\right]\right.\nonumber\\
&+&2\left[(\varphi_0-s_{2c}\varphi_{cos})^2 
{\overline {\rm H}}(m_{H^0},m_{G^\pm},m_{G^\pm})+
(\varphi_0+s_{2c}\varphi_{cos})^2
{\overline {\rm H}}(m_{H^0},m_{H^\pm},m_{H^\pm})\right]\nonumber\\
&+&4c_{2c}^2\left[\varphi_{sin}^2
{\overline {\rm H}}(m_{h^0},m_{G^\pm},m_{H^\pm})+
\varphi_{cos}^2 
{\overline {\rm H}}(m_{H^0},m_{G^\pm},m_{H^\pm})\right]\nonumber\\
&+&4\left[\varphi_{0,i}^2 
{\overline {\rm H}}(m_{G^0},m_{G^\pm},m_{H^\pm})+
\varphi_{v,i}^2 
{\overline {\rm H}}(m_{A^0},m_{G^\pm},m_{H^\pm})\right]\nonumber\\
&+&3c_{2r}^2\left[\varphi_{cos}^2
{\overline {\rm H}}(m_{h^0},m_{h^0},m_{h^0})+
\varphi_{sin}^2
{\overline {\rm H}}(m_{H^0},m_{H^0},m_{H^0})\right]\nonumber\\
&+&(2\varphi_0-3c_{2r}\varphi_{sin})^2
{\overline {\rm H}}(m_{H^0},m_{h^0},m_{h^0})
+(2\varphi_v-3c_{2r}\varphi_{cos})^2
{\overline {\rm H}}(m_{H^0},m_{H^0},m_{h^0})\nonumber\\
&+&c_{2i}^2\varphi_{cos}^2
\left[{\overline {\rm H}}(m_{h^0},m_{G^0},m_{G^0})
+{\overline {\rm H}}(m_{h^0},m_{A^0},m_{A^0})\right]\nonumber\\
&+&c_{2i}^2\varphi_{sin}^2
\left[{\overline {\rm H}}(m_{H^0},m_{G^0},m_{G^0})
+{\overline {\rm H}}(m_{H^0},m_{A^0},m_{A^0})\right]\nonumber\\
&+&\left.2s_{2i}^2\left[\varphi_{cos}^2
{\overline {\rm H}}(m_{h^0},m_{G^0},m_{A^0})+
\varphi_{sin}^2
{\overline {\rm H}}(m_{H^0},m_{G^0},m_{A^0})\right]\right\},\nonumber\\
V^{(p')}=&-&\frac{1}{2}N_c\left\{
[h_t^2c_r\varphi_2-g^2\varphi_{cos}/4]^2 
{\overline {\rm H}}(m_{h^0},\mtl,\mtl)+
(h_t^2c_r\varphi_2)^2{\overline {\rm H}}(m_{h^0},\mtr,\mtr)\right.\nonumber\\
&+&[h_t^2s_r\varphi_2+g^2\varphi_{sin}/4]^2
{\overline {\rm H}}(m_{H^0},\mtl,\mtl)+
(h_t^2s_r\varphi_2)^2{\overline {\rm H}}(m_{H^0},\mtr,\mtr)\nonumber\\
&+&
[h_t^2c_c\varphi_2+g^2\varphi_{sin,c}/2]^2
{\overline {\rm
H}}(m_{G^\pm},\mtl,\mbl)+
[g^2\varphi_{cos}/4]^2
{\overline {\rm H}}(m_{h^0},\mbl,\mbl)\nonumber\\
&+&\left.
[h_t^2s_c\varphi_2-g^2\varphi_{cos,c}/2]^2
{\overline {\rm H}}(m_{H^\pm},\mtl,\mbl)+
[g^2\varphi_{sin}/4]^2
{\overline {\rm H}}(m_{H^0},\mbl,\mbl)
\right\},\nonumber\\
V^{(z)}=&&\frac{1}{16}g^2\left\{
4c_{2c}^2\left[{\cal D}_{S}(m_{G^\pm},m_{G^\pm})+
{\cal D}_{S}(m_{H^\pm},m_{H^\pm})\right]
-4c_{4c}{\cal D}_{S}(m_{G^\pm},m_{H^\pm})\right.\nonumber\\
&+&2\left[ (1+s_{2c}s_{2r})\left[{\cal D}_{S}(m_{G^\pm},m_{h^0})+
{\cal D}_{S}(m_{H^\pm},m_{H^0})\right]\right.\nonumber\\
&+& (1-s_{2c}s_{2r})\left[{\cal D}_{S}(m_{G^\pm},m_{H^0})+
{\cal D}_{S}(m_{H^\pm},m_{h^0})\right]\nonumber\\
&+& (1-s_{2c}s_{2i})\left[{\cal D}_{S}(m_{G^\pm},m_{G^0})+
{\cal D}_{S}(m_{A^0},m_{H^\pm})\right]\nonumber\\
&+&\left. (1+s_{2c}s_{2i})\left[{\cal D}_{S}(m_{G^\pm},m_{A^0})+
{\cal D}_{S}(m_{G^0},m_{H^\pm})\right]\right]\nonumber\\
&+&\frac{3}{2}\left[c_{2r}^2\left[
{\cal D}_{S}(m_{h^0},m_{h^0})+{\cal D}_{S}(m_{H^0},m_{H^0})
\right]+c_{2i}^2\left[
{\cal D}_{S}(m_{G^0},m_{G^0})+{\cal D}_{S}(m_{A^0},m_{A^0})
\right]\right]\nonumber\\
&+&(2-3c_{2r}^2){\cal D}_{S}(m_{h^0},m_{H^0})+
(2-3c_{2i}^2){\cal D}_{S}(m_{G^0},m_{A^0})\nonumber\\
&+&\left.
c_{2i}c_{2r}\left[{\cal D}_{S}(m_{h^0},m_{G^0})+
{\cal D}_{S}(m_{H^0},m_{A^0})-{\cal D}_{S}(m_{h^0},m_{A^0})-
{\cal D}_{S}(m_{H^0},m_{G^0})\right]\right\},\nonumber\\
V^{(z')}=&&\frac{1}{2}N_ch_t^2
\left\{c_r^2[{\cal D}_{S}(\mtl,m_{h^0})+{\cal D}_{S}(\mtr,m_{h^0})]+
s_r^2[{\cal D}_{S}(\mtl,m_{H^0})+{\cal D}_{S}(\mtr,m_{H^0})]\right.
\nonumber\\
&+&
c_i^2[{\cal D}_{S}(\mtl,m_{G^0})+{\cal D}_{S}(\mtr,m_{G^0})]+
s_i^2[{\cal D}_{S}(\mtl,m_{A^0})+{\cal D}_{S}(\mtr,m_{A^0})]
\nonumber\\
&+&
\left. 2c_c^2[{\cal D}_{S}(\mtr,m_{G^\pm})+
{\cal D}_{S}(\mbl,m_{G^\pm})]+
2s_c^2[{\cal D}_{S}(\mtr,m_{H^\pm})+{\cal D}_{S}(\mbl,m_{H^\pm})]\right\}
\nonumber\\
&+&
\frac{1}{8}g^2N_c
\left\{ c_{2r}[{\cal D}_{S}(\mbl,m_{h^0})-{\cal D}_{S}(\mtl,m_{h^0})
+{\cal D}_{S}(\mtl,m_{H^0})-{\cal D}_{S}(\mbl,m_{H^0})]\right.\nonumber\\
&+&
c_{2i}[{\cal D}_{S}(\mtl,m_{A^0})-{\cal D}_{S}(\mbl,m_{A^0})
+{\cal D}_{S}(\mbl,m_{G^0})-{\cal D}_{S}(\mtl,m_{G^0})]\nonumber\\
&+&
\left.2c_{2c}[{\cal D}_{S}(\mbl,m_{H^\pm})-{\cal D}_{S}(\mtl,m_{H^\pm})
+{\cal D}_{S}(\mtl,m_{G^\pm})-{\cal D}_{S}(\mbl,m_{G^\pm})]\right\},
\nonumber\\
V^{(z'')}=&&
\frac{g^2}{4}N_c(2-N_c){\cal D}_{S}(\mtl,\mbl)
+h_t^2N_c\left[{\cal D}_{S}(\mtl,\mtr)+{\cal 
D}_{S}(\mbl,\mtr)\right]\nonumber\\
&+&\left(\frac{g^2}{8}+\frac{g_s^2}{6}\right)N_c(N_c+1)\left[
{\cal D}_{S}(\mtl,\mtl)+{\cal D}_{S}(\mbl,\mbl)
\right]\nonumber\\
&+&\frac{g_s^2}{6}N_c(N_c+1)\left[{\cal D}_{S}(\mtr,\mtr)+{\cal 
D}_{S}(\mbr,\mbr) \right]
\eear
where
$$
{\cal D}_{S}(m_1,m_2)=I(m_1)I(m_2). 
$$

In the previous formulae, dimensional regularization (with 
$n-1=3-2\epsilon$) is used to evaluate divergent integrals. Poles in 
$1/\epsilon$ and $\iota_\epsilon$-dependent terms cancel when 
counterterms are included. In addition, counterterms contribute a finite 
piece to the potential. 

With $g'=0$ the $T=0$ counterterm potential reads
\def\dzpu{\delta Z_{\varphi_1^2}}
\def\dzpd{\delta Z_{\varphi_2^2}}
\def\dzg{\delta Z_{g^2}}
\def\dzht{\delta Z_{h_t^2}}
\def\dzl{\delta Z_{\lambda}}
\bear
\nonumber
\delta V_{count}
&=&\frac{1}{2}\left\{
\frac{3}{4}(3-2\epsilon)I^\epsilon_\beta\left(\dzg+\dzpu+\dzpd\right)g^2
(\varphi_1^2+\varphi_2^2)
-6h_t^2I^\epsilon_{ f\beta}(\dzht+\dzpd)\varphi_2^2\right\},\nonumber
\eear
with 
\bear
\nonumber
I^\epsilon_\beta=\frac{T^2}{12}(1+\epsilon\iota_\epsilon)&,&
I^\epsilon_{ f\beta}=-\frac{T^2}{24}[1+\epsilon(\iota_\epsilon-\log4)].
\eear

The $\ov{\rm MS}$ renormalization functions 
$\delta Z$ (calculated in a 
model with two Higgs doublets plus third generation squarks) are
\[
\delta Z_{g^2}=-\frac{5}{2}g^2\frac{1}{16\pi^2\epsilon},\;\; \;\;
\delta Z_{\varphi_1^2}= \frac{9}{4}g^2\frac{1}{16\pi^2\epsilon},\;\; \;\;
\delta Z_{\varphi_2^2}= 
\left(\frac{9}{4}g^2-3h_t^2\right)\frac{1}{16\pi^2\epsilon},
\]
\[
\delta Z_{h_t^2}=\left(\frac{9}{2}h_t^2
-8g_s^2-\frac{9}{4}g^2\right)\frac{1}{16\pi^2\epsilon}.\;\; \;\;
\]

The finite contribution is then
\be
\label{count}
\delta 
V_{count,fin}=\frac{T^2}{64\pi^2}\left[
M^2(3h_t^2-2g^2)+m_t^2(16g_s^2-3h_t^2)\log 2\right].
\ee

The resummation procedure we followed for scalars can be implemented
by adding and subtracting the appropriate thermal mass terms for the scalars
in the Lagrangian. In this way the unperturbed Lagrangian contains already 
thermally corrected masses and thermal counterterms appear. To cancel all 
divergences, diagrams involving those counterterms should be included.
For the two--loop potential, one--loop thermal counterterm contributions 
are needed and give
\be
\label{thcount}
\delta V_{th,count}=-\frac{1}{2}\sum_{i,scalars}n_i\Pi_iI(m_i^2),
\ee
where $\Pi_i$ is the thermal self--energy of the $i^{th}$ scalar.

\vspace{0.5cm}
\section{ High T expansion of the Potential }

We give the dominant two--loop terms of the potential presented
in appendix B using a high T expansion.

First we have logarithmic terms coming from bosonic setting--sun
diagrams:
\bear
\delta V^{(a)}_{log} =  -\left. \frac{g^2}{64 \pi^2} T^2 \right\{
\cos^2(\theta_r+\theta_c) & \left[
\begin{array}{c}
 \\ \\
\end{array}
 \right. & 
(M^2-2m_{G^\pm}^2-2m_{h^0}^2) 
\log \left(\frac{3T}{m_{G^\pm}+m_{h^0}+M} \right)  
\nonumber \\
& + & \frac{(m_{G^\pm}^2-m_{h^0}^2)^2}{M^2} \log \left(
\frac{m_{G^\pm}+m_{h^0}}{m_{G^\pm}+m_{h^0}+M} \right) 
\nonumber \\
& + & (M^2-2m_{H^\pm}^2-2m_{H^0}^2) \log 
\left(\frac{3T}{m_{H^\pm}+m_{H^0}+M} \right) \nonumber \\
& + & \left. \frac{(m_{H^\pm}^2-m_{H^0}^2)^2}{M^2} \log \left(
\frac{m_{H^\pm}+m_{H^0}}{m_{H^\pm}+m_{H^0}+M} \right) \;\;\;\;\right]
\nonumber \\
+ \sin^2(\theta_r+\theta_c) & \left[ 
\begin{array}{c}
 \\ \\
\end{array}
\right. &
(M^2-2m_{G^\pm}^2-2m_{H^0}^2) \log 
\left(\frac{3T}{m_{G^\pm}+m_{H^0}+M} \right) \nonumber \\
& + & \frac{(m_{G^\pm}^2-m_{H^0}^2)^2}{M^2} \log \left(
\frac{m_{G^\pm}+m_{H^0}}{m_{G^\pm}+m_{H^0}+M} \right) 
\nonumber \\
& + & (M^2-2m_{H^\pm}^2-2m_{h^0}^2) \log
\left(\frac{3T}{m_{H^\pm}+m_{h^0}+M} \right) \nonumber \\ 
& + & \left. \frac{(m_{H^\pm}^2-m_{h^0}^2)^2}{M^2} \log \left(
\frac{m_{H^\pm}+m_{h^0}}{m_{H^\pm}+m_{h^0}+M} \right) \;\;\;\;\right] 
\nonumber \\
+ \cos^2(\theta_c-\theta_i) & \left[ 
\begin{array}{c}
 \\ \\
\end{array}
\right. &  
(M^2-2m_{G^\pm}^2-2m_{G^0}^2) 
\log \left(\frac{3T}{m_{G^\pm}+m_{G^0}+M} \right) \nonumber \\
& + & \frac{(m_{G^\pm}^2-m_{G^0}^2)^2}{M^2} \log \left(
\frac{m_{G^\pm}+m_{G^0}}{m_{G^\pm}+m_{G^0}+M} \right) 
\nonumber \\
& + & (M^2-2m_{H^\pm}^2-2m_{A^0}^2) \log
\left(\frac{3T}{m_{H^\pm}+m_{A^0}+M} \right) \nonumber \\
& + & \left. \frac{(m_{H^\pm}^2-m_{A^0}^2)^2}{M^2} \log \left(
\frac{m_{H^\pm}+m_{A^0}}{m_{H^\pm}+m_{A^0}+M} \right) \;\;\;\;\right] 
 \\
+ \sin^2(\theta_c-\theta_i) & \left[ 
\begin{array}{c}
 \\ \\
\end{array}
\right. &  
(M^2-2m_{G^\pm}^2-2m_{A^0}^2) \log 
\left(\frac{3T}{m_{G^\pm}+m_{A^0}+M} \right) \nonumber \\
& + & \frac{(m_{G^\pm}^2-m_{A^0}^2)^2}{M^2} \log \left(
\frac{m_{G^\pm}+m_{A^0}}{m_{G^\pm}+m_{A^0}+M} \right) 
\nonumber \\
& + & (M^2-2m_{H^\pm}^2-2m_{G^0}^2) \log
\left(\frac{3T}{m_{H^\pm}+m_{G^0}+M} \right) \nonumber \\
& + & \left. \frac{(m_{H^\pm}^2-m_{G^0}^2)^2}{M^2} \log \left(
\frac{m_{H^\pm}+m_{G^0}}{m_{H^\pm}+m_{G^0}+M} \right) \;\;\;\;\right]
\nonumber \\
+ \frac{1}{2} \cos^2(\theta_i+\theta_r) & \left[ 
\begin{array}{c}
 \\ \\
\end{array}
\right. & 
(M^2-2m_{G^0}^2-2m_{h^0}^2) \log 
\left(\frac{3T}{m_{G^0}+m_{h^0}+M} \right) \nonumber \\
& + & \frac{(m_{G^0}^2-m_{h^0}^2)^2}{M^2} \log \left(
\frac{m_{G^0}+m_{h^0}}{m_{G^0}+m_{h^0}+M} \right) 
\nonumber \\
& + & (M^2-2m_{A^0}^2-2m_{H^0}^2) \log
\left(\frac{3T}{m_{A^0}+m_{H^0}+M} \right) \nonumber \\
& + & \left. \frac{(m_{A^0}^2-m_{H^0}^2)^2}{M^2} \log \left(
\frac{m_{A^0}+m_{H^0}}{m_{A^0}+m_{H^0}+M} \right) \;\;\;\;
\right] \nonumber \\
+ \frac{1}{2} \sin^2(\theta_i+\theta_r) & \left[ 
\begin{array}{c}
 \\ \\
\end{array}
\right. & 
(M^2-2m_{G^0}^2-2m_{H^0}^2) \log 
\left(\frac{3T}{m_{G^0}+m_{H^0}+M} \right) \nonumber \\
& + & \frac{(m_{G^0}^2-m_{H^0}^2)^2}{M^2} \log \left(
\frac{m_{G^0}+m_{H^0}}{m_{G^0}+m_{H^0}+M} \right) 
\nonumber \\
& + & (M^2-2m_{A^0}^2-2m_{h^0}^2) \log
\left(\frac{3T}{m_{A^0}+m_{h^0}+M} \right) \nonumber \\
& + & \left. \frac{(m_{A^0}^2-m_{h^0}^2)^2}{M^2} \log \left(
\frac{m_{A^0}+m_{h^0}}{m_{A^0}+m_{h^0}+M} \right) \;\;\;\;\right] 
\nonumber \\
+ \frac{1}{2} (M^2-4m_{G^{\pm}}^2) \log \left( 
\frac{3T}{2m_{G^{\pm}}+M} \right) & + & \left. \frac{1}{2} 
(M^2-4m_{H^{\pm}}^2) \log \left( 
\frac{3T}{2m_{H^{\pm}}+M} \right) \right\} \nonumber  
\eear
\bear
\delta V^{(b)}_{log}  = -\frac{3g^4}{64 \pi^2} \left. \frac{T^2}{8} 
\right\{ \varphi_v^2 & \left[ 
\begin{array}{c}
 \\ \\
\end{array}
\right. & \frac{m_{h^0}^4}{2M^4} 
\left[ \log \left( \frac{m_{h^0}+M}{m_{h^0}+2M} \right)  
+ \log \left( \frac{m_{h^0}+M}{m_{h^0}} \right) \right]
\nonumber \\
+2\log\left(
\frac{m_{h^0}+2M}{m_{h^0}+2M_L}
\right)
& + & \left.  5 \log \left( \frac{3T}{m_{h^0}+2M} \right) + 
\frac{M^2-2m_{h^0}^2}{M^2} \log  \left( 
\frac{m_{h^0}+M}{m_{h^0}+2M} \right) \;\;\;\;\right] \nonumber\\
+ \varphi_0^2 & \left[ 
\begin{array}{c}
 \\ \\
\end{array}
\right. & \frac{m_{H^0}^4}{2M^4} 
\left[ \log \left( \frac{m_{H^0}+M}{m_{H^0}+2M} \right)
+ \log \left( \frac{m_{H^0}+M}{m_{H^0}} \right) \right] \\
+2\log\left(
\frac{m_{H^0}+2M}{m_{H^0}+2M_L}
\right)
& + & \left. \left. 5 \log \left( \frac{3T}{m_{H^0}+2M} \right) +
\frac{M^2-2m_{H^0}^2}{M^2} \log  \left( 
\frac{m_{H^0}+M}{m_{H^0}+2M} \right) \;\;\;\;\right] \right\} \nonumber
\eear
\be
\delta V^{(m+n)}_{log} = \frac{g^2}{64 \pi^2} T^2 \left[ 33 M^2
\log \left(\frac{T}{M} \right) - 6 (M^2-4M^2_L) \log \left(
\frac{3T}{2M_L+M} \right) \right]
\ee
\bear
\delta V^{(p)}_{log} & = & -\frac{g^4}{64 \pi^2} \frac{T^2}{16} 
\left\{ 3 c_{2r}^2 \left[ \varphi_{cos}^2 \log \left( 
\frac{T}{m_{h^0}} \right) + \varphi_{sin}^2 \log \left( 
\frac{T}{m_{H^0}} \right) \right]  \right. \nonumber \\
& + & 2(\varphi_v - s_{2c} \varphi_{sin})^2 \log \left( 
\frac{3T}{m_{h^0}+2m_{G^{\pm}}} \right) + 2(\varphi_0 - s_{2c} 
\varphi_{cos})^2 \log \left( 
\frac{3T}{m_{H^0}+2m_{G^{\pm}}} \right) \nonumber \\
& + & 2(\varphi_v + s_{2c} \varphi_{sin})^2 \log \left( 
\frac{3T}{m_{h^0}+2m_{H^{\pm}}} \right) + 2(\varphi_0 + s_{2c} 
\varphi_{cos})^2 \log \left( \frac{3T}{m_{H^0}+2m_{H^{\pm}}} 
\right) \nonumber \\
& + & 4 c_{2c}^2 \left[ \varphi_{sin}^2 \log \left( 
\frac{3T}{m_{h^0}+m_{G^{\pm}}+m_{H^{\pm}}} \right) + 
\varphi_{cos}^2 \log \left( 
\frac{3T}{m_{H^0}+m_{G^{\pm}}+m_{H^{\pm}}} \right) \right] 
\nonumber \\
& + & 4 \left[ \varphi_{0,i}^2 \log \left( 
\frac{3T}{m_{G^0}+m_{G^{\pm}}+m_{H^{\pm}}} \right) + 
\varphi_{v,i}^2 \log \left( 
\frac{3T}{m_{A^0}+m_{G^{\pm}}+m_{H^{\pm}}} \right) \right] 
\nonumber \\
& + & (2\varphi_0-3c_{2r} \varphi_{sin})^2 \log
\left( \frac{3T}{m_{H^0}+2m_{h^0}} \right) +  (2\varphi_v
-3c_{2r} \varphi_{cos})^2 \log \left( \frac{3T}{2m_{H^0}+m_{h^0}} 
\right) \nonumber \\
& + &  c_{2i}^2 \left( \varphi_{cos}^2 \left[ \log 
\left( \frac{3T}{m_{h^0}+2m_{G^0}} \right) +  \log \left( 
\frac{3T}{m_{h^0}+2m_{A^0}} \right) \right] \right. \nonumber \\
& + & \left. \varphi_{sin}^2 \left[ \log 
\left( \frac{3T}{m_{H^0}+2m_{G^0}} \right) +  \log \left( 
\frac{3T}{m_{H^0}+2m_{A^0}} \right) \right] \right) \nonumber \\
& + & \left. 2s_{2i}^2 \left[
\varphi_{cos}^2 \log \left( \frac{3T}{m_{h^0}+m_{G^0}+m_{A^0}} 
\right) + \varphi_{sin}^2 \log \left( 
\frac{3T}{m_{H^0}+m_{G^0}+m_{A^0}} \right) \right] \right\}  
\eear
and including squarks:
\bear
\delta V^{(a')}_{log} & = & -\frac{g^2}{128 \pi^2} T^2 N_c \left\{
4 (M^2-2m_{\tilde{t}_L}^2-2m_{\tilde{b}_L}^2) \log \left( 
\frac{3T}{m_{\tilde{t}_L}+m_{\tilde{b}_L}+M} \right) \right.
\nonumber \\
& + & 4 \frac{(m_{\tilde{t}_L}^2-m_{\tilde{b}_L}^2)^2}{M^2} 
\log \left( 
\frac{m_{\tilde{t}_L}+m_{\tilde{b}_L}}{m_{\tilde{t}_L}+m_{\tilde{b}_L}+M}
\right) \nonumber \\
& + & \left. (M^2-4m_{\tilde{t}_L}^2) \log \left( 
\frac{3T}{2m_{\tilde{t}_L}+M} \right) +  (M^2-4m_{\tilde{b}_L}^2) \log
\left( \frac{3T}{2m_{\tilde{b}_L}+M} \right) \right\} \nonumber \\
& + & \frac{g_s^2}{16 \pi^2} T^2 (N_c^2-1) \left\{ m_{\tilde{t}_L}^2 
\log \left( \frac{3T}{2m_{\tilde{t}_L}} \right) +  m_{\tilde{b}_L}^2 
\log \left( \frac{3T}{2m_{\tilde{b}_L}} \right) 
\right. \\
& + & \left. m_{\tilde{t}_R}^2 \log \left( \frac{3T}{2m_{\tilde{t}_R}}
\right) +  m_{\tilde{b}_R}^2 
\log \left( \frac{3T}{2m_{\tilde{b}_R}} \right) \right\} \nonumber
\eear
\bear
\delta V^{(p')}_{log} & = & -\frac{N_c}{32 \pi^2} T^2 \left\{
(h_t^2 c_r \varphi_2)^2 \log \left(
\frac{3T}{m_{h^0}+2m_{\tilde{t}_R}} \right) + (h_t^2 s_r \varphi_2)^2 
\log \left( \frac{3T}{m_{H^0}+2m_{\tilde{t}_R}} \right) 
 \right. \nonumber \\
& + & (h_t^2 c_r \varphi_2-\frac{g^2}{4} \varphi_{cos})^2 \log \left(
\frac{3T}{m_{h^0}+2m_{\tilde{t}_L}} \right) + (h_t^2 s_r \varphi_2+
\frac{g^2}{4} \varphi_{sin})^2 \log \left(
\frac{3T}{m_{H^0}+2m_{\tilde{t}_L}} \right) \nonumber \\
& + & (\frac{g^2}{4} \varphi_{cos})^2 \log \left(
\frac{3T}{m_{h^0}+2m_{\tilde{b}_L}} \right) + (\frac{g^2}{4} 
\varphi_{sin})^2 \log \left(
\frac{3T}{m_{H^0}+2m_{\tilde{b}_L}} \right) \nonumber \\
& + & (h_t^2 c_c \varphi_2- \frac{g^2}{2} \varphi_{cos,c})^2 \log 
\left( \frac{3T}{m_{G^{\pm}}+m_{\tilde{t}_L}+m_{\tilde{b}_L}} \right)
\nonumber \\
& + & \left. (h_t^2 s_c \varphi_2-\frac{g^2}{2} \varphi_{sin,c})^2 
\log \left(
\frac{3T}{m_{H^{\pm}}+m_{\tilde{t}_L}+m_{\tilde{b}_L}} \right) \right\} 
\eear

The linear terms needed to insure $\partial V/\partial \varphi_1=
\partial V/ \partial\varphi_2=0$ at $\varphi_1=\varphi_2=0$ are
\bear
\delta V^{(a)}_{lin}&=&-\frac{3}{8} g^2 M(m_{h^0}+2m_{G^\pm}+
m_{G^0}+ m_{H^0}+2m_{H^\pm}+m_{A^0}) \frac{T^2}{16\pi^2},
\eear
\bear
\delta V^{(b)}_{lin}&=&\frac{3}{64}g^4\frac{1}{M}(\varphi_v^2 m_{h^0}
+\varphi_0^2 m_{H^0})\frac{T^2}{8\pi^2},
\eear
\bear
\delta V^{(c)}_{lin}&=&-2\delta V^{(a)}_{lin},
\eear
\bear
\delta V^{(m+o)}_{lin}&=&3g^2MM_L\frac{T^2}{16\pi^2},
\eear
and
\bear
\delta V^{(a')}_{lin}&=&
-g^2N_c \frac{3M}{2}(\mstl+\msbl)\frac{T^2}{32\pi^2}\\
\delta V^{(c')}_{lin}&=&
3g^2N_c M(\mstl+\msbl)\frac{T^2}{32\pi^2}.
\eear
It can be easily checked that ($i=1,2$)
\[
\nonumber
\left.\frac{\partial }{\partial \varphi_i}\left(
\delta V^{(a)}_{log+lin}+\delta V^{(c)}_{lin}
\right)\right|_{\varphi=0}=0,
\]
a similar relation for $(a'+c')$, and
\[
\left.\frac{\partial }{\partial \varphi_i}\left(
\delta V^{(m+o)}_{log+lin}+\delta V^{(b+n)}_{log}
\right)\right|_{\varphi=0}=0.
\]

The non-logarithmic terms involving only $g_s$ and $h_t$ couplings are:

For Standard Model diagrams
\bear
\delta V^{(i+j)}&=&
\frac{m_t^2T^2}{64\pi^2}\left[
16 g_s^2 \left(\frac{5}{3}\log2-\frac{1}{2}-c_B\right)+
3N_c h_t^2 \left(\frac{5}{3}\log2-c_B\right)
\right],
\eear
with $c_B=\log(4\pi)-\gamma_E$, and from (\ref{count})
\be
\delta V_{count}=\frac{m_t^2T^2}{64\pi^2}(16g_s^2-3h_t^2)\log 2.
\ee
Diagrams involving squarks give
\bear
\delta V^{(a')}&=&-\frac{g_s^2T^2}{64\pi^2}(N_c^2-1)(c_2-1)(\mstl^2
+\mstr^2),\nonumber\\
\delta V^{(c')}&=&\frac{g_s^2T^2}{32\pi^2}(N_c^2-1)\left[
\Pi_{gL}^{1/2}(\mstl+\mstr)+\frac{1}{6}(\mstl^2+\mstr^2)
\right]+\delta_{c_B}V^{(c')},\nonumber\\
\delta V^{(p')}&=&\frac{3N_cT^2}{128\pi^2}c_2h_t^4\varphi_2^2,\nonumber\\
\delta V^{(z')}&=&\frac{N_cT^2}{32\pi^2}h_t^2\left[(\mstl+\mstr)(c_r^2m_{h^0}
+s_r^2m_{H^0}+s_i^2m_{A^0}+c_i^2m_{G^0})\right.\nonumber\\
&+&\left.2(\mstr+\msbl)(c_c^2m_{G^\pm}+s_c^2m_{H^\pm})\right]+
\delta_{c_B}V^{(z')},\nonumber\\
\delta V^{(z'')}&=&\frac{N_cT^2}{16\pi^2}\left[
\frac{g_s^2}{6}(N_c+1)(\mstl^2+\mstr^2)+h_t^2\mstr(\mstl+\msbl)
\right]+\delta_{c_B}V^{(z'')},
\eear
with $c_2\simeq 3.3025$.

The terms $\delta_{c_B} V$ add up to
$$
\delta_{c_B} V=\frac{c_B}{16\pi^2}\sum_{i,scalars}n_im_i^2\Pi_i.
$$
The contribution from thermal counterterms, eq.~(\ref{thcount}), gives 
exactly the same but with opposite sign so that they cancel out. If 
resummation of scalars is implemented only on zero Matsubara modes, 
there are no thermal counterterms. In that case $\delta_{c_B} V$ 
combines with the one-loop unresummed scalar contribution
\be
\frac{c_B}{32\pi^2}\sum_in_im_i^4,
\ee
to give
\be
\frac{c_B}{32\pi^2}\sum_in_i\overline{m}_i^4.
\ee
That is precisely the one-loop result if we resum all modes, so 
that after the cancellation of two--loop contributions and thermal 
counterterms, both resummation methods give the same result. This is 
no longer true in the presence of particle mixing (this occurs in our 
case in the Higgs sector) where a small numerical difference is expected.

\end{document}